\documentclass[a4paper,11pt]{article}
\pdfoutput=1 

\usepackage{jcappub} 

\usepackage[T1]{fontenc} 
\usepackage{verbatim}
\usepackage[utf8]{inputenc}
\usepackage[american]{babel}
\usepackage{graphicx}
\usepackage{epsfig}
\usepackage{adjustbox}
\usepackage{booktabs}
\usepackage{multirow}
\usepackage{dcolumn}
\usepackage{amsmath}
\usepackage{amsfonts}
\usepackage{mathtools}
\usepackage{amssymb}
\usepackage{subcaption} 
\usepackage{caption}
\usepackage{epstopdf}
\usepackage{bm}
\usepackage{mathrsfs}
\usepackage{version}
\usepackage{enumerate}
\usepackage{braket}
\usepackage{enumitem}
\usepackage{transparent}
\usepackage{pifont}
\usepackage{colortbl}
\usepackage{rotating}
\usepackage{adjustbox}
\usepackage{cancel}
\usepackage{tabularx}
\usepackage{soul}
\usepackage[table]{xcolor}
\usepackage{siunitx}
\usepackage{float} 

\usepackage{hyperref}
\hypersetup{colorlinks=true,
linkcolor=navyblue,citecolor=navyblue,urlcolor=navyblue,pdfencoding=auto}

\definecolor{navyblue}{rgb}{0.0, 0.0, 0.5}
\definecolor{royalblue}{rgb}{0.25, 0.41, 0.88}
\definecolor{cadmiumgreen}{rgb}{0.0, 0.42, 0.24}
\definecolor{blue-violet}{rgb}{0.54, 0.17, 0.89}
\definecolor{darkviolet}{rgb}{0.58, 0.0, 0.83}
\definecolor{orange(colorwheel)}{rgb}{1.0, 0.5, 0.0}

\definecolor{magenta(process)}{rgb}{1.0, 0.0, 0.56}

\definecolor{darkspringgreen}{rgb}{0.09, 0.45, 0.27}

\definecolor{royalblue(web)}{rgb}{0.25, 0.41, 0.88}

\definecolor{cadmiumorange}{rgb}{0.93, 0.53, 0.18}

\definecolor{heliotrope}{rgb}{0.87, 0.45, 1.0}

\makeatletter
\renewcommand*{\@textcolor}[3]{%
\protect\leavevmode
\begingroup
\color#1{#2}#3%
\endgroup
}
\makeatother

\newcommand{\myfloatalign}{\centering}

\renewcommand\[{\left[}

\DeclarePairedDelimiter{\abs}{\lvert}{\rvert}

\makeatletter
\let\save@mathaccent\mathaccent
\newcommand*\if@single[3]{%
\setbox0\hbox{${\mathaccent"0362{#1}}^H$}%
\setbox2\hbox{${\mathaccent"0362{\kern0pt#1}}^H$}%
\ifdim\ht0=\ht2 #3\else #2\fi
}
\newcommand*\rel@kern[1]{\kern#1\dimexpr\macc@kerna}
\newcommand*\widebar[1]{\@ifnextchar^{{\wide@bar{#1}{0}}}{\wide@bar{#1}{1}}}
\newcommand*\wide@bar[2]{\if@single{#1}{\wide@bar@{#1}{#2}{1}}{\wide@bar@{#1}{#2}{2}}}
\newcommand*\wide@bar@[3]{%
\begingroup
\def\mathaccent##1##2{%
\let\mathaccent\save@mathaccent
\if#32 \let\macc@nucleus\first@char \fi
\setbox\z@\hbox{$\macc@style{\macc@nucleus}_{}$}%
\setbox\tw@\hbox{$\macc@style{\macc@nucleus}{}_{}$}%
\dimen@\wd\tw@
\advance\dimen@-\wd\z@
\divide\dimen@ 3
\@tempdima\wd\tw@
\advance\@tempdima-\scriptspace
\divide\@tempdima 10
\advance\dimen@-\@tempdima
\ifdim\dimen@>\z@ \dimen@0pt\fi
\rel@kern{0.6}\kern-\dimen@
\if#31
\overline{\rel@kern{-0.6}\kern\dimen@\macc@nucleus\rel@kern{0.4}\kern\dimen@}%
\advance\dimen@0.4\dimexpr\macc@kerna
\let\final@kern#2%
\ifdim\dimen@<\z@ \let\final@kern1\fi
\if\final@kern1 \kern-\dimen@\fi
\else
\overline{\rel@kern{-0.6}\kern\dimen@#1}%
\fi
}%
\macc@depth\@ne
\let\math@bgroup\@empty \let\math@egroup\macc@set@skewchar
\mathsurround\z@ \frozen@everymath{\mathgroup\macc@group\relax}%
\macc@set@skewchar\relax
\let\mathaccentV\macc@nested@a
\if#31
\macc@nested@a\relax111{#1}%
\else
\def\gobble@till@marker##1\endmarker{}%
\futurelet\first@char\gobble@till@marker#1\endmarker
\ifcat\noexpand\first@char A\else
\def\first@char{}%
\fi
\macc@nested@a\relax111{\first@char}%
\fi
\endgroup
}
\makeatother

\newcommand\ee{\end{equation}}
\newcommand\be{\begin{equation}}
\newcommand\eea{\end{eqnarray}}
\newcommand\bea{\begin{eqnarray}}
\newcommand{\bsp}{\begin{split}}
\newcommand{\esp}{\end{split}}
\newcommand{\bit}{\begin{itemize}[leftmargin=*]}
\newcommand{\eit}{\end{itemize}}
\newcommand{\ben}{\begin{enumerate}[leftmargin=*]}
\newcommand{\een}{\end{enumerate}}

\newcommand{\ie}{\textit{i.e.}~}
\newcommand{\eg}{\textit{e.g.}~}

\newcommand{\mpc}[1]{\SI{#1}{\mathrm{Mpc}^{-1}}}

\newcommand\eq[1]{Eq.~\eqref{eq:#1}}

\newcommand{\eqsII}[2]{Eqs.~\eqref{eq:#1}, \eqref{eq:#2}}
\newcommand{\eqsIII}[3]{Eqs.~\eqref{eq:#1}, \eqref{eq:#2}, \eqref{eq:#3}}

\newcommand{\dif}{\mathrm{d}}

\renewcommand{\vec}{\bm} 
\newcommand\vers[1]{\hat{\vec{#1}}}

\newcommand\eps{\varepsilon}

\newcommand{\prt}[2]{\frac{\partial #1}{\partial #2}}

\newcommand{\cH}{\mathcal{H}}

\def\mpl{M_{\rm P}}
\def\ns{n_\mathrm{s}}
\def\pic{\pi_\text{c}}

\def\fnleq{f_\mathrm{NL}^\mathrm{equil}}
\def\fnlloc{f_\mathrm{NL}^\mathrm{loc}}
\def\fnlres{f_\mathrm{NL}^\mathrm{res}}

\DeclareMathOperator{\sech}{sech}

\title{Imprints of Oscillatory Bispectra on Galaxy Clustering}

\author[a]{G. Cabass,}
\author[b]{E. Pajer,}
\author[a]{F. Schmidt}

\affiliation[a]{Max-Planck-Institut f\"{u}r Astrophysik, 
Karl-Schwarzschild-Str. 1, 85741 Garching, Germany}
\affiliation[b]{Institute for Theoretical Physics and Center for Extreme Matter and Emergent Phenomena,
	Utrecht University, 
	Princetonplein 5, 3584 CC Utrecht, The Netherlands}

\emailAdd{gcabass@mpa-garching.mpg.de}
\emailAdd{e.pajer@uu.nl}
\emailAdd{fabians@mpa-garching.mpg.de}

\abstract{\noindent Long-short mode coupling during inflation, encoded in the squeezed bispectrum of curvature perturbations, 
induces a dependence of the local, small-scale power spectrum on long-wavelength perturbations, leading to a scale-dependent halo bias. 
While this scale dependence is absent in the large-scale limit for single-field inflation models that satisfy the consistency relation, 
certain models such as resonant non-Gaussianity show a peculiar behavior on intermediate scales. 
We reconsider the predictions for the halo bias in this model by working in Conformal Fermi Coordinates, 
which isolate the physical effects of long-wavelength perturbations on short-scale physics. 
We find that the bias oscillates with scale with an envelope similar to that of equilateral non-Gaussianity. 
Moreover, the bias shows a peculiar modulation with the halo mass. 
Unfortunately, we find that upcoming surveys will be unable to detect the signal because of its very small amplitude. 
We also discuss non-Gaussianity due to interactions between the inflaton and massive fields: 
our results for the bias agree with those in the literature. }

\begin{document}
\maketitle
\flushbottom


\section{Introduction}
\label{sec:intro}

\noindent One of the main goals of modern cosmology is understanding the dynamics of inflation. 
Special interest goes to the interactions among the degrees of freedom active during inflation. 
These will affect the correlation functions of primordial fluctuations and lead to a deviation from Gaussianity: 
since the same fluctuations provide the initial conditions for structure formation, 
we expect to see an imprint of primordial interactions in the higher-order correlation functions of late-time observables.
Current constraints on non-Gaussianity come mainly from the CMB \cite{Ade:2015ava}, 
but significant improvements are expected to come from large-scale structure (LSS) observations 
\cite{Alvarez:2014vva,Baldauf:2016sjb,Welling:2016dng,Gleyzes:2016tdh}. 

The scale dependence of the bias of tracers of the dark matter density field (\eg dark matter halos) 
is a prominent signature of primordial non-Gaussianity on large-scale structure 
\cite{Matarrese:2008nc,Slosar:2008hx,Desjacques:2010jw,Desjacques:2016bnm}. 
For Gaussian initial conditions, the number density of halos on scales 
larger than the halo Lagrangian radius $R_\ast$, \ie in a region of size $R_L\gg R_\ast$, 
depends on the average of the matter density in that region (and its gradients). 
There is also dependence on the amplitude of matter fluctuations on 
scales smaller than $R_L^{-1}$: 
however, in absence of a coupling between long- and short-wavelength modes in the initial conditions, this dependence only affects 
the tracer statistics on large scales by adding a white-noise term to the otherwise deterministic relation between their number density and the matter density. 
Things change for non-Gaussian initial conditions. 
In this case there are large-scale modulations of the small-scale fluctuations, due to mode coupling. 
This leads, in general, to a dependence on both the amplitude and the shape of the small-scale matter power spectrum: 
both are correlated with long-wavelength perturbations, in a way that is encoded by the squeezed-limit bispectrum. 
Famously, for local primordial non-Gaussianity, where the primordial bispectrum takes the form 
$B_\phi(\vec{k}_1,\vec{k}_2,\vec{k}_3)=2f_\mathrm{NL}^\mathrm{loc}\big(P_\phi(k_1)P_\phi(k_2)+\text{2 perms.}\big)$, 
the amplitude of small-scale fluctuations is uniformly rescaled by the gravitational potential, 
leading to a scale-dependent bias that increases as $k^{-2}$ on large scales relative to matter \cite{Dalal:2007cu}. 
Looking for this scale dependence in galaxy surveys 
can place constraints on non-Gaussianity that are competitive with those from the CMB: 
future large-volume galaxy surveys can in principle reach the precision of $\sigma(f_\mathrm{NL}^\mathrm{loc})\lesssim\num{1}$ 
\cite{Ferramacho:2014pua,Yamauchi:2014ioa,Ferraro:2014jba,Raccanelli:2014awa,dePutter:2014lna,Castorina:2018zfk}, 
and can go down to $\sigma(f_\mathrm{NL}^\mathrm{loc})\approx\num{0.2}$ when measurements of the galaxy bispectrum are included \cite{Karagiannis:2018jdt}. 

While sizable local non-Gaussianity is generically expected in multi-field inflationary models, things are different in single-clock inflation. 
Here, the leading behavior of the bispectrum in the squeezed limit is determined by the so-called consistency relation \cite{Maldacena:2002vr,Creminelli:2004yq}, 
which relates it to the tilt of the power spectrum, giving $f_\mathrm{NL}^\text{loc} = 1-n_\text{s}$. 
However, such consistency relation cancels from any local observable \cite{Pajer:2013ana,Dai:2015rda,Dai:2015jaa,dePutter:2015vga}: 
when discussing dark matter tracers this is just the statement that, with adiabatic initial conditions, their clustering 
is not sensitive to the local scale factor, but can be affected only by the local curvature and the local Hubble rate 
\cite{Baldauf:2011bh,Wagner:2014aka,Baldauf:2015vio,Desjacques:2016bnm}. 
This ``separate universe'' picture is made explicit by working in Conformal Fermi Coordinates (CFC) \cite{Dai:2015rda,Dai:2015jaa}: 
in CFC the squeezed bispectrum contains only the physical part of the couplings between long and short modes set up by inflation, 
which can actually affect the dynamics once the modes re-enter the Hubble radius.

Consider, then, the squeezed-limit expansion in powers of $k_\ell/k_s$, where $k_\ell^{-1}$ and $k_s^{-1}$ are, 
respectively, the scale at which we measure correlations and the size of the tracers we are looking at. 
If the consistency relation holds, using CFC one can show that the first physical effect from the initial conditions arises at order $(k_\ell/k_s)^2$ \cite{Pajer:2013ana}. 
For some particular inflationary models, however, this conclusion is too quick. 
Consider for example the case of an inflaton speed of sound $c^2_\mathrm{s}\ll 1$: 
in this case, there is an enhancement $\sim c^{-2}_\mathrm{s}(k_\ell/k_s)^2$ in the squeezed limit \cite{Creminelli:2013cga}. 
Another example is that of axion monodromy-inspired models of inflation \cite{Silverstein:2008sg,McAllister:2008hb,Flauger:2009ab}, 
where the slow-roll potential possesses small sinusoidal modulations. 
In these models, the bispectrum has a specific shape (called resonant non-Gaussianity) 
which has little overlap with all other bispectrum templates \cite{Flauger:2010ja}. 
The bispectrum satisfies the consistency relation \cite{Flauger:2010ja}, 
but deviations appear at order $\alpha k_\ell/k_s$, 
where $\alpha\gg 1$ is the frequency of the oscillations of the background in units of the Hubble rate \cite{Creminelli:2011rh,Behbahani:2011it,Pajer:2013ana}. 
This is due to the fact that in the regime of rapid oscillations, 
the modifications in the power spectrum come from resonant production of inflaton quanta of comoving momentum $k$ happening at $\frac{k}{aH}\sim\alpha$, 
\ie when the mode is a factor $\alpha$ shorter than the Hubble radius. 
For the long mode to act as a modification of the background for the short modes, 
it had to be well outside the Hubble radius not only at the time of Hubble exit of the short modes, 
but already when they were in resonance \cite{Flauger:2010ja,Creminelli:2011rh}. 
A third example is Khronon inflation \cite{Creminelli:2012xb}: 
in this case the time dependence of the curvature perturbation $\zeta(k)$ does not start at order $k^2$ as in the standard scenario, 
so there is a non-trivial contribution at order $k_\ell$ in the squeezed limit, 
which corresponds to a non-local term $\propto\sqrt{\partial^2} \zeta$. 
However, this contribution will be suppressed by the deviation from scale invariance, 
\ie by slow-roll, so the final result is too small to be of interest. 

A second family of models that show a peculiar behavior in the squeezed limit are the so-called ``cosmological collider'' models. 
Here, the inflaton is coupled to other massive fields that can have non-zero spin: 
the conversion of a long-wavelength massive particle into curvature perturbations can 
generate long-short mode couplings and leave an imprint on the squeezed bispectrum 
\cite{Chen:2009zp,Chen:2012ge,Arkani-Hamed:2015bza,Lee:2016vti}. 
While the consistency relation is satisfied, \ie the term $\propto(k_\ell/k_s)^0$ is fixed by $1-\ns$, 
it is possible to have a non-analytical dependence of the scalar bispectrum on $k_\ell/k_s$ at sub-leading order in the squeezed limit: 
indeed, while some of the effects of the interaction can be mimicked by adding 
local vertices to the Lagrangian of the Goldstone boson $\pi$ of broken time diffeomorphisms 
(as a result of integrating out these massive fields), 
spontaneous particle production in an expanding spacetime will lead to effects that are inherently non-local. 
As an example, consider a massive scalar field $\sigma$ coupled to $\pi$: 
if $m\geq 3H/2$ the bispectrum will show an oscillatory behavior with $k_\ell/k_s$, 
while fields with $0<m<3H/2$ will lead to scalings $(k_\ell/k_s)^{3/2-\nu}$, where $\nu = \sqrt{9/4-m^2/H^2}$. 
While these non-analytical scalings are suppressed by $\exp({-{\pi m}/{H}})$, 
it is possible to obtain some observable effect in the limit of small sound speed for $\pi$ \cite{Lee:2016vti}. 
Unlike isocurvature modes, 
the energy density due to massive particles becomes quickly diluted during inflation, 
so that they do not have any effects on observables after the end of inflation. 
In other words, 
the relationship between the adiabatic mode $\pi$ at the end of inflation 
and the conserved quantity $\zeta$ is the same as in the single-clock case. 
This allows us, again, to study the effect of this primordial non-Gaussianity on the scale-dependent bias 
by working in Conformal Fermi Coordinates, 
\ie by computing the squeezed CFC bispectrum. 

One last class of single-field models known as Ultra Slow-Roll inflation \cite{Kinney:2005vj} 
is well-known for its peculiar features in the squeezed-limit bispectrum \cite{Namjoo:2012aa,Chen:2013aj,Akhshik:2015nfa}. 
New soft theorems in this class of models have been recently discussed in \cite{Mooij:2015yka,Finelli:2017fml,Bravo:2017wyw,Finelli:2018upr}, 
together with their implications for late-universe observables \cite{Bravo:2017gct}. Here, we will not discuss them further. 

The goal of this paper is to compute the scale-dependent bias in these inflationary models. 
Among the single-clock models discussed above, 
we are going to focus only on the case of resonant non-Gaussianity: 
this was originally investigated in \cite{CyrRacine:2011rx} 
(which, however, included the unphysical contribution from the consistency relation). 
For the cosmological collider models, we consider only the case of a spin-$0$ field with $m\geq 3H/2$ 
(the more general case was recently investigated in \cite{MoradinezhadDizgah:2017szk}). 

The paper is organized as follows: 
in Section \ref{sec:bias_and_curvature} we review the concept of bias in General Relativity following \cite{Baldauf:2011bh}, 
matching to the results of \cite{Dai:2015jaa}: 
the technical details of how to use the calculation of CFC bispectrum during inflation 
as the initial condition for the late-time evolution are collected in Appendix \ref{app:appendix_A}. 
In Section \ref{sec:resonant_intro} we compute the CFC bispectrum for resonant non-Gaussianity, 
and discuss the results for the bias in Section \ref{sec:bias_resonant_NG} 
(some theoretical details on the calculation of the bias are collected in Appendix \ref{app:resummation_proof}, 
while two additional plots are shown in Appendix \ref{app:appendix_B}). 
Section \ref{sec:QSF_intro} does the same for cosmological collider models. We conclude in Section \ref{sec:conclusions}. 
Appendix \ref{app:appendix_D} contains the details of the calculation of the CFC bispectrum in cosmological collider models. 

\paragraph{Notation and conventions}
We follow \cite{Maldacena:2002vr,Baldauf:2011bh,Wagner:2014aka} and denote the comoving curvature perturbation by $\zeta$. 
In \cite{Dai:2015jaa} it is denoted by $\cal{R}$: more precisely $\zeta_\text{here} = -\cal{R}_\text{there}$. We denote conformal time by $\eta$. 
The Planck mass is defined as $\mpl=(8\pi G_{\rm N})^{-1/2}$. The amplitude of the dimensionless scalar power spectrum is defined as 
$\mathcal{A}_\text{s} = \Delta^2_\zeta$.

\section{Bias as dependence from local curvature}
\label{sec:bias_and_curvature}

\noindent We briefly review the definition of halo bias from \cite{Baldauf:2011bh}, 
and see how it can be matched to the calculations of \cite{Dai:2015jaa,Cabass:2016cgp}. 
The formation of dark matter halos happens on scales of order of the non-linear scale 
(the distance travelled by dark matter particles): 
this is the short scale $k_s^{-1}$. 
Consider then an adiabatic perturbation in the gravitational potentials of wavelength $k_\ell^{-1}$ much longer than the size of the region that forms a halo. 
If the mode is also longer than the sound horizon of short-scale perturbations, 
its effect will be equivalent to that of a modification of the local expansion history and of the local curvature of the universe. 

Then, halos will form as if they were living in a different background FLRW, 
and we can define bias as how a family of inertial observers living in the patch 
sees the number density of halos depend on the mean curvature that they measure in the patch.\footnote{This must then be connected to 
what is actually measured by an observer on Earth. 
However, these ``projection effects'' are independent of the dynamics, and can be computed separately, for example 
with the so-called ruler perturbations of \cite{Schmidt:2012ne,Jeong:2013psa,Jeong:2014ufa} 
(see also \cite{Bertacca:2014wga,Kehagias:2015tda,DiDio:2016gpd} for similar approaches).} 
In Conformal Fermi Coordinates $(t_F,\vec{x}_F = \vec{q})$, 
where $t_F$ is the proper time and $\vec{q}$ is the Lagrangian coordinate of comoving observers with the matter fluid, 
this ``separate universe picture'' is manifest, and we can define the Eulerian \mbox{bias as} 
\begin{equation}
\label{eq:bias_CFC}
b_{K_F}^E \equiv -\prt{\log\braket{n_h^E(t_F,\vec{q})}}{\Omega_{K_F}}\bigg|_{K_F=0}\,\,,\qquad\Omega_{K_F}=-\frac{K_F}{a^2_FH^2_F}\,\,,
\end{equation}
\ie as the derivative of the proper density of halos of a given mass, averaged in the CFC patch, with respect to 
the contribution of the local curvature to the energy density. 
The definition that is usually used in the literature is different from \eq{bias_CFC}: 
indeed one defines the bias $b_1^E$ as a derivative with respect to the local density using the relation between $K_F$ and the synchronous-gauge 
density perturbation, \ie 
\begin{equation}
\label{eq:K_F_vs_b_1}
K_F=\frac{5}{3}\frac{\Omega_\mathrm{m}H^2_0}{D_1(\eta)}\delta_\text{sc}(\eta)\,\,,
\end{equation}
where $D_1(\eta)$ is the linear growth factor normalized to $a(\eta)$ during matter domination. 
In the rest of this paper we are going to use the Lagrangian bias $b_1$ as bias parameter, 
\emph{i.e.}\footnote{The relation between the Lagrangian and Eulerian biases 
is determined by 
\begin{equation*}
1+\delta_h^E=(1+\delta_h)(1+\delta_\text{sc})=1+(1+b_1)\delta_\text{sc}\,\,,
\end{equation*}
where $\delta_h$ and $\delta_h^E$ are the fractional Lagrangian and Eulerian halo density perturbations at fixed proper time, respectively. } 
\begin{equation}
\label{eq:b_1}
b_1 = b_1^E-1=\frac{5}{3}\frac{\Omega_\mathrm{m}H^2_0}{D_1(\eta)}\frac{b_{K_F}}{a^2_FH^2_F}-1\,\,.
\end{equation}

Non-linear gravitational evolution will contribute to the coupling of the short-scale modes to local curvature 
(mainly through the modification to the expansion history if the speed of sound is small), 
thereby contributing to the derivative in \eq{bias_CFC} and then to $b_1$. 
Therefore, if the initial small-scale perturbations are not correlated with the long-wavelength mode, 
$b_1$ captures the full dependence of the halo number density on it (at leading order in derivatives and in perturbations). 
However, some contribution will also come from the initial conditions, 
\ie by the long-short mode coupling during inflation. 
Here CFC are again very useful. 
Indeed, we can use them to follow both short and long modes from when they leave the Hubble radius 
during inflation to Hubble re-entry during radiation or matter dominance 
(this would not have been possible had we used standard Fermi Normal Coordinates, 
since they can be trusted only far inside the Hubble scale): 
more precisely, it is enough to work at linear order in perturbation theory for the short modes in the curved separate universe described by CFC. 
The initial conditions for the Newtonian potentials, then, are simply given by 
\begin{equation}
\label{eq:initial_conditions}
\phi_s^F = \psi_s^F = -\frac{3(1+w)}{5+3w}\zeta_s^F\,\,,
\end{equation}
where we have defined the potentials $\phi_s^F$ and $\psi_s^F$ in \textit{Newtonian gauge} by 
\begin{equation}
\label{eq:newtonian_gauge}
\dif s^2 = a^2_F\bigg[{-(1+2\phi^F_s)\dif\eta^2_F} + \frac{(1-2\psi^F_s)\dif\vec{x}_F^2}{(1 + K_F\abs{\vec{x}_F}^2/4)^2}\bigg]\,\,.
\end{equation}
The procedure is delineated in Appendix \ref{app:appendix_A}: 
it amounts to computing the relation between the gauge-invariant curvature perturbation on constant energy hypersurfaces in a curved FLRW background 
and the short-scale metric perturbations in Newtonian gauge. 
We can already understand why \eq{initial_conditions} will hold by recalling that the corrections coming 
from curvature will scale as $\Omega_{K_F}/H^2_F = -K_F/(a^2_F H^2_F)$, 
which is exponentially diluted during inflation. 
All that is left, then, is how interactions during inflation couple $\zeta_s^F$ to $K_F$. 
These are the initial conditions that one can use, \emph{e.g.}, in a separate universe simulation: 
for example, one can run numerical codes such as CAMB \cite{Lewis:1999bs} with non-zero curvature and modified expansion history 
and an initial power spectrum for the comoving curvature perturbation equal to the one in CFC 
\cite{Baldauf:2011bh,Wagner:2014aka,Baldauf:2015vio}.\footnote{Of course, 
to be able to treat $K_F$ as a constant throughout, the long mode must re-enter the Hubble radius during matter dominance.} 

What about the coupling with $K_F$ or, equivalently, the squeezed bispectrum of the comoving curvature perturbation in CFC? 
This can be derived in a number of ways \cite{Creminelli:2013cga,Cabass:2016cgp}, 
but the most straightforward way is to realize that, 
if we are in presence of an isotropic long mode $\zeta_\ell$, 
the long-wavelength part of the metric during inflation is already of the same form of that of a curved FLRW, 
apart from corrections that decay as the universe expands. 
Indeed, let us consider \textit{comoving gauge} and neglect tensor modes: 
at linear order in perturbations, the metric takes the form
\begin{equation}
\label{eq:maldacena_gauge-A}
\begin{split}
&\dif s^2 = a^2(\eta)\big[{-\big(1+2 {\delta N_\ell}(\eta,\vec{x})\big)\dif\eta^2} \\
&\hphantom{\dif s^2 = a^2(\eta)\big[ } + N_{i,\ell}(\eta,\vec{x})(\dif\eta\dif x^i + \dif x^i\dif\eta) \\
&\hphantom{\dif s^2 = a^2(\eta)\big[ } + \big(1+2\zeta_\ell(\eta,\vec{x})\big)\dif\vec{x}^2\big] + \mathcal{O}(\zeta_\ell^2)\,\,.
\end{split}
\end{equation}
At this order, using the isotropy of the long mode (\ie $\partial_i\zeta_\ell(\eta,\vec{0}) = 0$ and 
$\partial_i\partial_j\zeta_\ell(\eta,\vec{0}) = \partial^2\zeta_\ell(\eta,\vec{0})\delta_{ij}/3$), 
we can rewrite the spatial part as
\begin{equation}
\label{eq:maldacena_gauge-B}
g_{ij} = \frac{a^2(\eta)\big[1+2\zeta_\ell(\eta,\vec{0})\big]\delta_{ij}}{\Big(1 + \frac{K_F\abs{\vec{x}}^2}{4}\Big)^2} + \mathcal{O}(\vec{x}^3,\zeta_\ell^2)\,\,,
\end{equation}
where $K_F = -2\partial^2\zeta_\ell(\eta,\vec{0})/3$. The solutions for the lapse and shift constraints, \ie \cite{Maldacena:2002vr}
\begin{subequations}
\label{eq:maldacena_gauge-C}
\begin{align}
&\delta N_\ell = \frac{\zeta'_\ell}{\cH}\,\,, \label{eq:maldacena_gauge-C-1} \\
&N_{i,\ell} = \partial_i\bigg({-\frac{\zeta_\ell}{\cH}} + \eps\partial^{-2}\zeta'_\ell\bigg)\,\,, \label{eq:maldacena_gauge-C-2}
\end{align}
\end{subequations}
decay at late times (indeed, since we assume that the vacuum is the Bunch-Davies one and that the background has already reached the attractor, 
the time dependence of $\zeta$ starts at order $k^2$: $\zeta'\sim k^2\eta$). Notice that here by ``late times'' we mean the transition from the inflationary phase to the Hot Big Bang phase. 
To find the initial conditions on super-Hubble scales for the evolution of perturbations in the epoch of radiation or matter dominance one takes the $\eta\to 0$ limit during 
the inflationary phase, when the expansion is close to de Sitter, $\mathcal{H} = -1/\eta$ 
(see \cite{Maldacena:2002vr} and \cite{Boubekeur:2008kn} for a calculation up to second order in 
perturbations). To summarize, this tells us that the CFC scale factor will asymptote to its unperturbed value at late times \cite{Creminelli:2013cga,Cabass:2016cgp}, 
and only the correction to the local curvature $K_F$, which is conserved on super-horizon scales, will survive.\footnote{As shown in \cite{Cabass:2016cgp}, 
there are $\mathcal{O}(\eps)$ tidal fields, that in any case vanish for an isotropic long mode.} 

Therefore, in the case of an isotropic long mode, the CFC calculation simplifies greatly, and reduces to that of \cite{Pajer:2013ana}: 
more precisely, up to and including $\mathcal{O}(k^2_\ell/k^2_s)$, 
the super-Hubble squeezed CFC bispectrum takes the form 
\begin{equation}
\label{eq:squeezed_CFC_bispectrum}
\begin{split}
B_{\zeta}^F(k_\ell,k_s) &= B_{\zeta}(k_\ell,k_s) - \frac{\dif\log\big(k_s^3P_\zeta(k_s)\big)}{\dif\log k_s}P_\zeta(k_\ell)P_\zeta(k_s) \\
&\equiv B_{\zeta}(k_\ell,k_s) - \Delta B_{\zeta}(k_\ell,k_s)\,\,,
\end{split}
\end{equation}
where $B_{\zeta}(k_\ell,k_s)$ is the squeezed bispectrum in global coordinates, 
averaged over the angle $\cos\theta$ between $\vec{k}_\ell$ and $\vec{k}_s$:
\begin{equation}
\int_{-1}^{1}\frac{\dif\cos\theta}{2}\lim_{\vec{k}_\ell\rightarrow 0}\braket{\zeta(\vec{k}_{1})\zeta(\vec{k}_2)\zeta(\vec{k}_{\ell})}
\equiv(2\pi)^{3}\delta^{(3)}(\vec{k}_1+\vec{k}_2+\vec{k}_\ell)B_{\zeta}(k_\ell,k_s)\,\,,
\end{equation}
with $\vec{k}_1=\vec{k}_s-\vec{k}/2$, $\vec{k}_2={-\vec{k}_s}-\vec{k}/2$. 
Since the bias is sensitive only to the angle-average of the squeezed-limit bispectrum, 
\eq{squeezed_CFC_bispectrum} is all we need to compute the contribution to $b_1$ from the initial conditions. 
We start by considering the case of resonant non-Gaussianity.

\section{Resonant non-Gaussianity}
\label{sec:resonant_intro}

\noindent The bispectrum of the curvature perturbation in axion monodromy-inspired models of inflation, 
in which a periodic term is superimposed on the slow-roll potential as
\begin{equation}
\label{eq:resonant_potential}
V(\phi) = V_0(\phi) + \Lambda^4\cos\bigg(\frac{\phi}{f}\bigg)\,\,,
\end{equation}
is given by \cite{Flauger:2010ja}
\begin{equation}
\label{eq:resonant_bispectrum}
\begin{split}
&B_\zeta(k_1,k_2,k_3) = \fnlres\frac{\big(4\pi^2\Delta^2_\zeta\big)^2}{\prod_{i=1}^3k_i^2}\bigg[\sin\bigg(\alpha\log\frac{k_t}{k_\ast}\bigg) 
+ \alpha^{-1}\sum_{i\neq j}\frac{k_i}{k_j}\cos\bigg(\alpha\log\frac{k_t}{k_\ast}\bigg) \\
&\hphantom{B_\zeta(k_1,k_2,k_3) = \fnlres\frac{\big(4\pi^2\Delta^2_\zeta\big)^2}{\prod_{i=1}^3k_i^2}\bigg[ } 
- \alpha^{-2}\frac{k_t\sum_{i=1}^3k^2_i}{\prod_{i=1}^3k_i}\sin\bigg(\alpha\log\frac{k_t}{k_\ast}\bigg)\bigg]+B_{\zeta}^{\text{ho}}(k_1,k_2,k_3)\,\,,
\end{split}
\end{equation}
where $k_t\equiv\sum_{i=1}^3 k_i$ and $k_\ast$ is the pivot scale at which the amplitude $\Delta^2_\zeta$ of the dimensionless power spectrum is defined. 
Here $B_{\zeta}^{\text{ho}}$ represents terms of higher order in the $\alpha\gg 1$ expansion, where $\alpha$ is related to the axion decay constant $f$ 
and the slow-roll parameter $\eps = -\dot{H}_\mathrm{sr}/H_\mathrm{sr}^2$ through $f/\sqrt{2\eps} = \mpl/\alpha$, 
and it can be related to the frequency $\omega$ of the periodic features in the Hubble rate 
$H(t)=H_\mathrm{sr}(t)+H_\mathrm{osc}(t)\sin(\omega t)$ by $\alpha = \omega/H_\mathrm{sr}$. 
The power spectrum, instead, is given by \cite{Flauger:2009ab,Meerburg:2010ks,Behbahani:2011it,Ade:2015lrj}
\begin{equation}
\label{eq:power_spectrum_resonant}
P_\zeta(k)=\frac{2\pi^2\Delta^2_\zeta}{k^3}\bigg(\frac{k}{k_\ast}\bigg)^{n_\text{s}-1}\bigg[1+\delta n_\text{s}\cos\bigg(\frac{\phi_k}{f}\bigg)\bigg]\,\,,
\end{equation}
where $\phi_k=\phi_\ast - \sqrt{2\eps_\ast}\log(k/k_\ast)$ is the value of the inflaton field at which the mode with comoving momentum $k$ exits the Hubble radius, 
and $\eps_\ast$ is the value of $\eps = -\dot{H}_\mathrm{sr}/H_\mathrm{sr}^2$ when the pivot scale $k_\ast$ exits the Hubble radius. 
Here, the amplitude of the sinusoidal correction to the scalar tilt $\delta n_\text{s}$ and the amplitude of non-Gaussianity $\fnlres$ are related by 
\begin{equation}
\label{eq:fNLres}
\fnlres = \frac{3b_\ast\sqrt{2\pi}}{8}\alpha^{\frac{3}{2}} = \frac{\alpha^2\delta n_\text{s}}{8}\,\,,
\end{equation}
where the monotonicity parameter $b_\ast$ quantifies the height of the modulations to the slow-roll potential 
(and must satisfy $b_\ast<1$ to avoid that the inflaton gets trapped in a local minimum). 

As discussed in \cite{Creminelli:2011rh,Pajer:2013ana,Cabass:2016cgp}, 
the squeezed bispectrum $B_{\zeta}$ should not contain terms proportional to $k_\ell/k_s$, 
if $\vec{k}_s$ and $\vec{k}_\ell$ are defined as 
\begin{equation}
\label{eq:three_k}
\vec{k}_1=\vec{k}_s-\frac{\vec{k}_\ell}{2}\,\,,\qquad\vec{k}_2={-\vec{k}_s}-\frac{\vec{k}_\ell}{2}\,\,,\qquad\vec{k}_3=\vec{k}_\ell\,\,.
\end{equation}
Instead, such linear terms do appear when expanding the three leading order terms in \eq{resonant_bispectrum}. 
The reason for their presence is the following. 
The bispectrum in global coordinates is computed as an expansion in $\alpha^{-1}$, 
and at a fixed order $\alpha^{-n}$, the term proportional to $\alpha^{-n}\times(k_\ell/k_s)$ in the squeezed limit will be cancelled by\footnote{This 
is seen explicitly in \eq{resonant_bispectrum}: since the explicit terms go up to order $\alpha^{-2}$, there are no terms $\sim\alpha^{-1}\times(k_\ell/k_s)$ in 
the squeezed global bispectrum (and then in the CFC transformation). }
a contribution coming from the term of order $\alpha^{-(n+1)}$ \cite{Creminelli:2011rh}. 
So, if we do not include $B_{\zeta}^{\text{ho}}$ we do not find a cancellation at order $\alpha^{-2}$. 
To fix this problem, we keep a contribution from the higher order term that 
in the squeezed limit takes precisely the right form to cancel all $k_\ell/k_s$ terms at order $\alpha^{-2}$: 
\begin{equation}
\label{eq:higher_order_bispectrum_term}
B_{\zeta}^{\text{ho}}(k_\ell,k_s)\supset {-\fnlres}\frac{8}{\alpha^2}\frac{k_\ell}{k_s}\sin\bigg(\alpha\log\frac{2 k_s}{k_\ast}\bigg)P_\zeta(k_\ell)P_\zeta(k_s)\,\,.
\end{equation}
By including this term but neglecting the corresponding term of order $\alpha^{-3}$ in the bispectrum in global coordinates of \eq{resonant_bispectrum}, 
we are making an error of order $\alpha^{-3}$, 
which is negligible for any foreseeable application (as we will see in Section \ref{sec:resonant_results}). 
The CFC correction $\Delta B_{\zeta}$ is simply computed from the power spectrum of \eq{power_spectrum_resonant}:
\begin{equation}
\label{eq:resonant_CFC_transformation}
\Delta B_\zeta(k_\ell,k_s) \equiv\fnlres\bigg\{\frac{8}{\alpha}\bigg[\cos\bigg(\alpha\log\frac{2 k_s}{k_\ast}\bigg) 
- \frac{2}{\alpha}\sin\bigg(\alpha\log\frac{2 k_s}{k_\ast}\bigg)\bigg] \bigg\}P_\zeta(k_\ell)P_\zeta(k_s)\,\,.
\end{equation}
In summary, the CFC squeezed bispectrum is given by 
\begin{equation}
\label{eq:enhancement}
\boxed{
\begin{split}
B^{F}_{\zeta} &= B_{\zeta} + B^{\text{ho}}_{\zeta} - \Delta B _{\zeta} \\
&=\fnlres \bigg[2\alpha\cos\bigg(\alpha\log\frac{2 k_s}{k_\ast}\bigg)
- \frac{10}{3}\sin\bigg(\alpha\log\frac{2 k_s}{k_\ast}\bigg) 
+ \dots\bigg]\frac{k^2_\ell}{k^2_s} P_\zeta(k_\ell)P_\zeta(k_s) + \mathcal{O}\bigg(\frac{k^3_\ell}{k^3_s}\bigg)\,\,.
\end{split}
}
\end{equation}
Some comments are in order:
\begin{itemize}
\item subleading corrections in $\alpha^{-1}$ to the bispectrum of \eq{resonant_bispectrum} will correct 
\eq{enhancement} only at order $\alpha^{-3}\times(k_\ell^2/k_s^2)$; 
\item at a fixed value of $k_\ell/k_s$, 
we see that the ``physical'' term of order $k_\ell^2/k_s^2$ is enhanced for large $\alpha$;
\item at order $(k_\ell/k_s)^0$, the term out of phase with the tilt of the power spectrum is 
cancelled by the contribution of order $\alpha^{-3}$ to \eq{resonant_bispectrum}.
\end{itemize}

Having derived the CFC bispectrum, in the next section we briefly discuss what is the maximum value of $\fnlres$ currently allowed by data. 
Then, in Section \ref{sec:bias_resonant_NG} we discuss how the correction to $b_1$ from general primordial non-Gaussianity 
can be computed in practice, and show the results for the resonant case.

\subsection{Large frequency limit and maximum \texorpdfstring{$f_\mathrm{NL}^\mathrm{res}$}{f\^{}\{res\}\_\{NL\}}}
\label{sec:large_alpha}

\noindent Let us review the theoretical and observational priors on the frequency of oscillations $\alpha$. 
As explained in \cite{Flauger:2009ab,Flauger:2010ja}, 
the theoretically motivated regime is $\alpha\gg1$: 
indeed, as we discussed above, $\alpha$ is equal to $\sqrt{2\varepsilon}\,\mpl/f = 0.1(\mpl/f)$, 
(fixing $\eps$ to the maximum value allowed by current bounds on the tensor-to-scalar ratio, 
\ie $r<0.08$ at $95\%\,\mathrm{CL}$ \cite{Ade:2015lrj,Array:2015xqh}) 
where $f$ is the axion decay constant. 
String theory seems to predict sub-Planckian decay constants \cite{Banks:2003sx,Svrcek:2006yi}, 
and $f\ll\mpl$ is possible. 
The $\alpha\gg 1$ limit is interesting also from the point of view of the observational prospects, 
since we recall that $\fnlres$ is given by \eq{fNLres}, \ie $\fnlres = {3b_\ast\sqrt{2\pi}}\alpha^{\frac{3}{2}}/{8} = {\alpha^2\delta n_\text{s}}/{8}$. 

\begin{figure}[H]
\myfloatalign
\centering
\includegraphics[width=0.7\columnwidth]{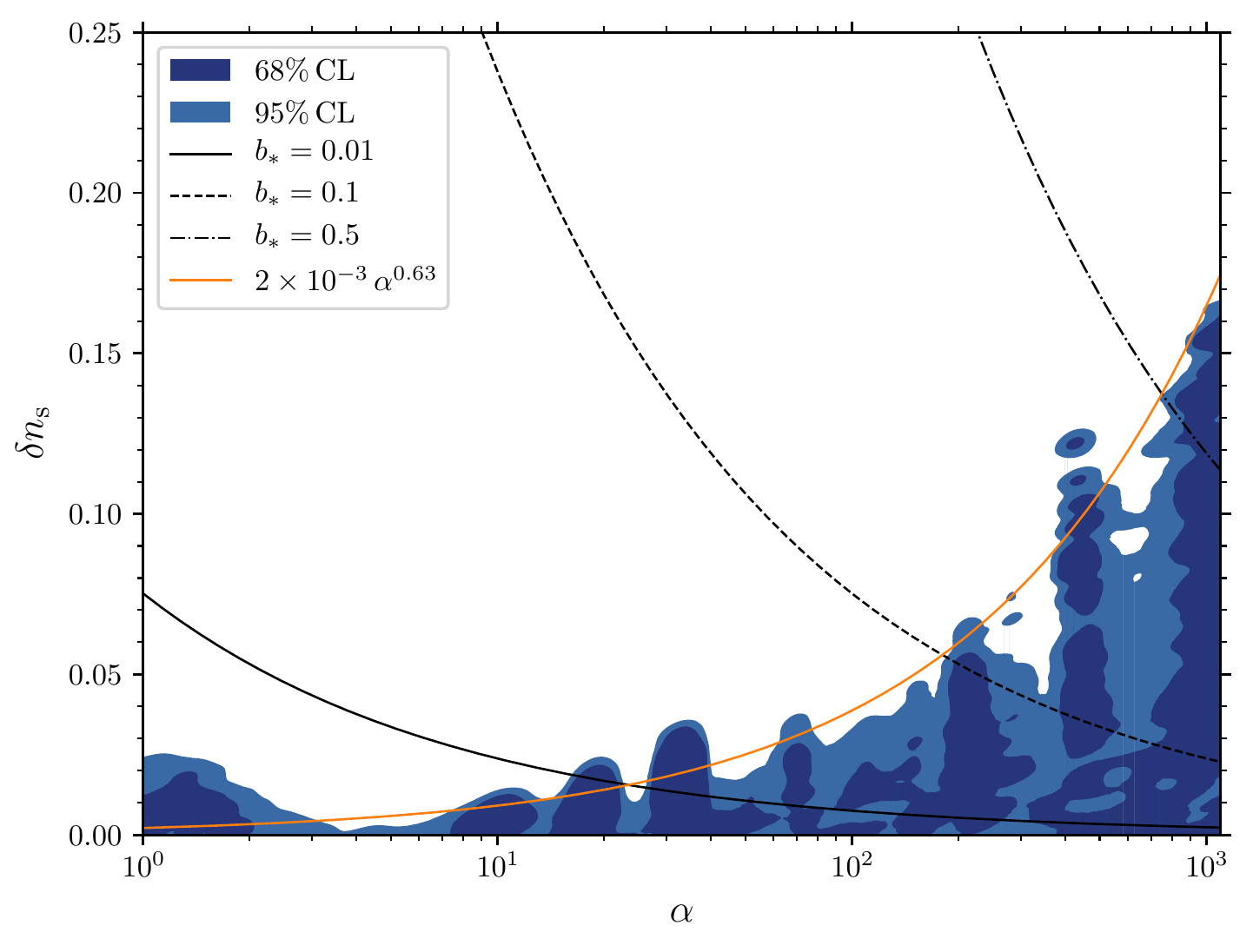} 
\caption{Constraints in the $\alpha$ -- $\delta n_\text{s}$ plane from \emph{Planck} temperature and polarization angular spectra. 
We see that the whole $95\%\,\mathrm{CL}$ allowed region is consistent with having a monotonicity parameter $b_\ast < 1$. 
Therefore, we can safely take $\delta n_\text{s} = \num{2d-3}\times\alpha^{0.63}$ (orange line) 
as a rough estimate for the maximum value of $\delta n_\text{s}$ allowed by \emph{Planck}. }
\label{fig:fig_40_planck}
\end{figure}

We start, then, by discussing what is the maximum value of $\alpha$ 
(and, with it, of $\fnlres$) 
that we can achieve given the current bounds from CMB anisotropies. 
Current $95\%\,\mathrm{CL}$ bounds from \emph{Planck} temperature and polarization angular spectra, 
taken from Fig.~40 of \cite{Ade:2015lrj}, are reproduced in Fig.~\ref{fig:fig_40_planck} 
(we reprocess the figure with \emph{Mathematica} to extract the 
$68\%\,\mathrm{CL}$ and $95\%\,\mathrm{CL}$ contours).\footnote{A 
forecast on how much future CMB and LSS experiments will improve the bounds on a primordial power spectrum with superimposed logarithmic oscillations 
has been recently carried out in \cite{Ballardini:2016hpi}. } 
From Fig.~\ref{fig:fig_40_planck} we see that a good estimate for the maximum $\fnlres$ allowed at $95\%\,\mathrm{CL}$ 
is $\fnlres\approx\num{3d-4}\times\alpha^{\num{2.63}}$, going up to $\fnlres\approx\num{2.6d4}$ for $\alpha = e^7\approx\num{1.1d3}$ 
(\ie the maximum value considered in the \emph{Planck} analysis). 

There are additional constraints, both theoretical and observational, that we must mention. 
One constraint on the maximum value of $\alpha$ comes from the requirement of perturbative non-Gaussianity. 
For primordial fluctuations of size $\braket{\zeta^2}\approx\num{2d-9}$, we see that 
a perturbative treatment of their PDF around a Gaussian breaks down if the three-point function 
$\braket{\zeta^3}\sim\fnlres\braket{\zeta^2}^{2}$ becomes of order $\braket{\zeta^2}^{3/2}$, requiring $\fnlres\lesssim\num{2d4}$. 
Using \eq{fNLres}, this translates into the upper bound $\alpha\lesssim\num{800}\times b_\ast^{-2/3}$. 
From Fig.~\ref{fig:fig_40_planck} it is straightforward to see that, in the range of frequencies considered by \emph{Planck}, 
the requirement of perturbative non-Gaussianity is satisfied. 
Another constraint comes from the CMB: 
with it we can place bounds on resonant non-Gaussianity directly through, \emph{e.g.}, the $TTT$ bispectrum. 
In \cite{Ade:2015ava}, the scan for resonant non-Gaussianity in the bispectrum has been carried out for $\log\alpha < \num{3.9}$, 
with the conclusion that the constraints on $\fnlres$ are a factor of $\approx 8$ weaker than the ones coming from the power spectrum. 
Finally, in \cite{Behbahani:2011it}, 
resonant non-Gaussianity was studied in the context of the Effective Field Theory of Inflation, 
by studying the case where the continuous shift symmetry of the Goldstone boson of time diffeomorphisms is softly broken to a discrete subgroup. 
Bounds on the consistency of the Effective Field Theory, then, imply a constraint 
$\alpha\ll (4\pi/\sqrt{2\braket{\zeta^2}})^{1/2}\approx\num{400}$. 
It is important to stress, however, that this bound is obtained from knowledge of the low-energy theory only: 
if, \emph{e.g.}, the UV completion in string theory is known, 
it would be replaced by (model-dependent) requirements of consistency and computability 
in the string compactifications that give rise to a potential like the one of \eq{resonant_potential} 
(we refer to \cite{Flauger:2009ab} for a discussion). 
For these reasons, 
in the following we will carry out a phenomenological analysis, 
and consider as constraints on $\alpha$ and $b_\ast$ only those coming from current data.

\section{Bias in resonant non-Gaussianity}
\label{sec:bias_resonant_NG}

\noindent The effect of primordial non-Gaussianity on the statistics of dark matter halos can be treated in a model-independent way through the bias expansion, 
that describes the halo density contrast $\delta_h$ as a local 
functional of gravitational observables (like $\delta_\text{m}$ or the tidal field 
$K_{ij} = (\partial_i\partial_j/\partial^2-\delta_{ij}/3)\delta_\text{m}$).\footnote{Strictly speaking, $\delta_\text{m}$ is an observable only on sub-Hubble scales. 
Moreover, we stress that the dependence of $\delta_h$ on $\delta_\text{m}$ and $K_{ij}$ is local in space but non-local in time. 
For non-Gaussian initial conditions, the time evolution induces higher-order operators that are important for the galaxy bispectrum \cite{Assassi:2015fma}. } 
More precisely, since we are interested in the clustering of halos 
on scales much larger than the typical spatial scales involved in their formation (of order the Lagrangian radius for halos), 
the kinematic regime of the three-point function that is relevant for the bias is the squeezed limit, 
where one mode $k_3 = k_\ell$ is much longer than the other two modes $k_1\sim k_2\sim k_s$. 

Let us expand the bispectrum of the Newtonian potential $\phi$ in this limit. 
Following \cite{Desjacques:2016bnm}, we can write (using \eq{three_k} as the definition of $\vec{k}_i$ in terms of $\vec{k}_s$, $\vec{k}_\ell$)
\begin{equation}
\label{eq:bias_exp-A}
\begin{split}
&B_\phi(\vec{k}_1,\vec{k}_2,\vec{k}_\ell) = A(\vec{k}_s,\vec{k}_\ell)P_\phi(k_s)P_\phi(k_\ell) + \mathcal{O}(k^2_\ell/k^2_s) \\
&\hphantom{B_\phi(\vec{k}_1,\vec{k}_2,\vec{k}_\ell) } = 
\sum_{J=0,2,\dots}A_{J}(k_s,k_\ell)\mathcal{L}_{J}(\cos\theta)P_\phi(k_s)P_\phi(k_\ell) + \mathcal{O}(k^2_\ell/k^2_s)\,\,,
\end{split}
\end{equation}
where $\mathcal{L}_J$ are the Legendre polynomials and we have defined $\cos\theta = \vers{k}_s\cdot\vers{k}_\ell$. 
Notice that only even Legendre polynomials can appear in the squeezed-limit expansion: 
indeed, we can interpret the squeezed bispectrum as a modulation by a long-wavelength mode $\phi(\vec{k}_\ell)$ of the 
local power spectrum $P_\phi(\vec{k}_s|\vec{q})$ in a patch much smaller than the wavelength of this mode (that we center in $\vec{q}=0$ for simplicity). 
Since this is the power spectrum of a real field, we must have $P_\phi(\vec{k}_s|\vec{q}) = P_\phi({-\vec{k}_s}|\vec{q})$. 

Let us now assume for a moment that $A_J(k_s,k_\ell) = 4a_J(k_\ell/k_s)^\Delta$ (the factor of $4$ is conventional), 
and focus on the $J=0$ case (since we are interested in the bias, which is sensitive only to the monopole of the bispectrum). 
\eq{bias_exp-A} tells us that the local power spectrum of the short-scale perturbations in this patch is modulated as
\begin{equation}
\label{eq:bias_exp-B}
P_\phi(\vec{k}_s|\vec{q}) = \bigg[1+4a_0\phi(\vec{k}_\ell)\bigg(\frac{k_\ell}{k_s}\bigg)^\Delta e^{i\vec{k}_\ell\cdot\vec{q}}\bigg]P_\phi(k_s)\,\,.
\end{equation}
For $\Delta=0$, we see that the amplitude of the small-scale power spectrum is uniformly enhanced by the long mode, 
while for $\Delta\neq 0$ we see that also its shape is affected. 
Since the modulation of $P_\phi(k_s)$ is mediated by $\Psi^{(\Delta)}(\vec{k}_\ell)\equiv k_\ell^\Delta\phi(\vec{k}_\ell)$, we see that 
at leading order in the squeezed limit we need to include in the bias expansion the following operator, 
written in terms of the Lagrangian coordinate $\vec{q}$: 
\begin{equation}
\label{eq:bias_exp-C}
\Psi^{(\Delta)}(\vec{q}) = \int\frac{\dif^3k}{(2\pi)^3}k^\Delta\phi(\vec{k})e^{i\vec{k}\cdot\vec{q}}\,\,.
\end{equation}

If we want to go beyond the squeezed limit, we need to take the expansion of \eq{bias_exp-A} to higher order, \ie (focusing on the $J=0$ case)
\begin{equation}
\label{eq:bias_exp-D}
\begin{split}
&A_0(k_s,k_\ell) = 4\sum_{n=0}^{+\infty}a_{0,2n}\bigg(\frac{k_\ell}{k_s}\bigg)^{\Delta+2n}\,\,,
\end{split}
\end{equation}
where we let the sum run over even powers of $k_\ell/k_s$, 
since odd powers would correspond to non-local couplings like, \emph{e.g.}, $\delta_h\sim\sqrt{\partial^2}\zeta$ for $\Delta = 0$ and $n=1/2$. 
From this we see that we need to add the operators $\Psi^{(\Delta+2)},\Psi^{(\Delta+4)},\dots$ to the bias expansion, 
with the corresponding bias coefficients $b_{\Psi^{(\Delta+2n)}}$. 
Moreover, spatial derivatives of these operators will be present as well, \ie we need to include operators like $\sim\partial^{2+2m}\Psi^{(\Delta+2n)}$. 

Let us now go back to the squeezed limit $n=0$. \eq{bias_exp-B} suggests a way to derive the actual values of the bias coefficient $b_{\Psi^{(\Delta)}}$: 
we need to include the modulation of the short modes induced by the squeezed bispectrum in the initial small-scale density fluctuations, 
follow the halo formation, and then compute the response of the halo number density to variations in $a_{0,0}\Psi^{(\Delta)}$. 
As we see from \eq{bias_exp-B}, for some infinitesimal $a_{0,0}\Psi^{(\Delta)} = \epsilon$ the rescaling of the initial $\delta_\text{m}(\vec{k})$ takes the form
\begin{equation}
\label{eq:bias_exp-E}
\delta_\text{m}(\vec{k})\to\big[1+2\epsilon k^{-\Delta}\big]\delta_\text{m}(\vec{k})\,\,.
\end{equation}
Then, calling $\widebar{n}_h$ the average density of halos inside the patch over which the long mode is slowly-varying 
(from now on we will follow the notation of \cite{Desjacques:2016bnm}), the bias coefficient $b_{\Psi^{(\Delta)}}$ is given by 
\begin{equation}
\label{eq:bias_exp-F}
b_{\Psi^{(\Delta)}} = \frac{\partial\log\widebar{n}_h}{\partial\epsilon}\bigg|_{\epsilon=0}\,\,.
\end{equation}
In the case of local-type non-Gaussianity $(\Delta=0)$, this was explicitly verified using simulations by \cite{Baldauf:2015vio,Biagetti:2016ywx}. 
One can also straightforwardly extend this to higher orders in the squeezed limit, \ie compute $b_{\Psi^{(\Delta+2n)}}$, 
by sending $\Delta\to\Delta + 2n$ in \eq{bias_exp-E}. 
Let us now assume that in the Gaussian case the number density of halos depends on the small-scale power spectrum through its variance on some scale $R$, 
that we take to be the Lagrangian radius $R_\ast = R_\ast(M)$ (where $M$ is the halo mass). 
Then, by dimensional analysis, we can expect that $b_{\Psi^{(\Delta+2n)}}\sim R_\ast^{\Delta+2n}$. 
This can be seen explicitly if we consider the case of a universal mass function, \ie
\begin{equation}
\label{eq:bias_exp-G}
\widebar{n}_h = \widebar{n}_h(\widebar{\rho}_\text{m},\sigma_\ast)\abs[\bigg]{\frac{\dif\log\sigma_\ast}{\dif\log M}}\,\,,
\end{equation}
where $\widebar{\rho}_\text{m}$ is the average matter density in the patch, and the variance over $R_\ast$ is defined as
\begin{equation}
\label{eq:bias_exp-H}
\sigma^2_\ast=\int\frac{\dif^3k}{(2\pi)^3}W^2_\ast(k)P_\text{m}(k)\,\,,
\end{equation}
with $W_\ast(k) = 3j_1(kR_\ast)/(kR_\ast)$ being the Fourier transform of a spherical top-hat filter with radius $R_\ast$. 
Indeed, fixing for a moment $n=0$, 
we can use the fact that under the transformation of \eq{bias_exp-E} the variance and the Jacobian $J=\abs{\dif\log\sigma_\ast/\dif\log M}$ transform as 
\begin{subequations}
\label{eq:bias_exp-I}
\begin{align}
&\sigma^2_{\ast}\to \bigg[1+2\epsilon\frac{\sigma^2_{\ast,-\Delta/2}}{\sigma^2_{\ast}}\bigg]\sigma^2_{\ast}\,\,, \label{eq:bias_exp-I-1} \\
&J\to \bigg[1+4\epsilon\frac{\sigma^2_{\ast,-\Delta/2}}{\sigma^2_{\ast}}\bigg(
\frac{\dif\log\sigma^2_{\ast,-\Delta/2}}{\dif\log\sigma^2_{\ast}}-1\bigg)\bigg]J\,\,, \label{eq:bias_exp-I-2}
\end{align}
\end{subequations}
where the generalized spectral moment $\sigma^2_{\ast,p}$ is defined as
\begin{equation}
\label{eq:bias_exp-J}
\sigma^2_{\ast,p}\equiv\int\frac{\dif^3k}{(2\pi)^3}k^{2p}W^2_\ast(k)P_\text{m}(k)\,\,.
\end{equation}
Then, \eq{bias_exp-F} becomes
\begin{equation}
\label{eq:bias_exp-K}
b_{\Psi^{(\Delta)}} = \bigg[b_{\Psi^{(0)}} + 4\bigg(\frac{\dif\log\sigma^2_{\ast,-\Delta/2}}{\dif\log\sigma^2_{\ast}}-1\bigg)\bigg]
\frac{\sigma^2_{\ast,-\Delta/2}}{\sigma^2_{\ast}}\,\,,
\end{equation}
where $b_{\Psi^{(0)}}=b_\phi$ is the bias parameter quantifying the effect of local primordial non-Gaus{\-}sia{\-}ni{\-}ty. 
The same procedure can be generalized to include orders beyond the squeezed limit as well, again by sending $\Delta\to\Delta + 2n$. 
Therefore, since the generalized spectral moment will scale as $(R_\ast)^{\Delta+2n}$, 
we find $b_{\Psi^{(\Delta+2n)}}\sim R_\ast^{\Delta+2n}$. 

This result can be further specialized to the case of a mass function of the form 
\begin{equation}
\label{eq:bias_exp-L}
\widebar{n}_h=\frac{\widebar{\rho}_\text{m}}{M}\,\nu_\text{c}f(\nu_\text{c})\,J\,\,,
\end{equation}
where $f$ is an arbitrary function of the significance $\nu_\text{c}\equiv\delta_\text{c}/\sigma_\ast$ 
(with $\delta_\text{c}$ being the threshold for spherical collapse).\footnote{The significance quantifies 
how rare fluctuations above $(1+\delta_\text{c})$ are, 
given a RMS amplitude $\sigma_\ast$.}
Indeed, we can now compute both $b_1$ and $b_\phi$. 
As discussed in Section \ref{sec:bias_and_curvature}, $b_1$ can be computed by taking a derivative with respect to $\widebar{\rho}_\text{m}$, 
\emph{i.e.}\footnote{Physically, 
there is a non-zero derivative because of the shifting of the threshold for spherical collapse due to the long mode. 
In the separate universe with density $\rho_\text{m}=\widebar{\rho}_\text{m}(1+\delta_\ell)$, 
where $\delta_\ell$ (called $\Delta$ in Section 3 of \cite{Desjacques:2016bnm}) 
is approximately uniform over the patch, a spherical perturbation will collapse when the total energy density becomes equal to 
$\rho_\text{c} = \rho_\text{m}(1+\delta_\text{c}) = \widebar{\rho}_\text{m}(1+\delta_\text{c}+\delta_\ell)$. 
An observer that can collect information from multiple separate universes, instead, 
would associate to the different regions different collapse thresholds $\widebar{\delta}_\text{c}$ 
such that $\rho_\text{c}=\widebar{\rho}_\text{m}(1+\widebar{\delta}_\text{c})$. 
From the two expressions for $\rho_\text{c}$ one gets $\delta_\text{c}=\widebar{\delta}_\text{c}-\delta_\ell$. 
That is, in an overdense region the threshold for spherical collapse is lowered, 
while an underdensity makes it larger. }
\begin{equation}
\label{eq:bias_exp-M}
\begin{split}
&b_1^E=1+b_1 = \frac{\partial\log\widebar{n}_h}{\partial\log\widebar{\rho}_\text{m}} = 
1-{\frac{1}{\sigma_\ast}}\frac{\dif\log\big[\nu_\text{c}f(\nu_\text{c})\big]}{\dif\nu_\text{c}}\,\,. 
\end{split}
\end{equation}
Using \eq{bias_exp-F}, instead, $b_\phi$ becomes
\begin{equation}
\label{eq:bias_exp-N}
b_\phi = \frac{\partial\log\widebar{n}_h}{\partial\epsilon} = 2\frac{\partial\log\widebar{n}_h}{\partial\log\sigma_\ast} = 2\delta_\text{c}b_1\,\,.
\end{equation}

\begin{figure}[H]
\myfloatalign
\centering
\begin{tabular}{c}
\includegraphics[width=0.65\columnwidth]{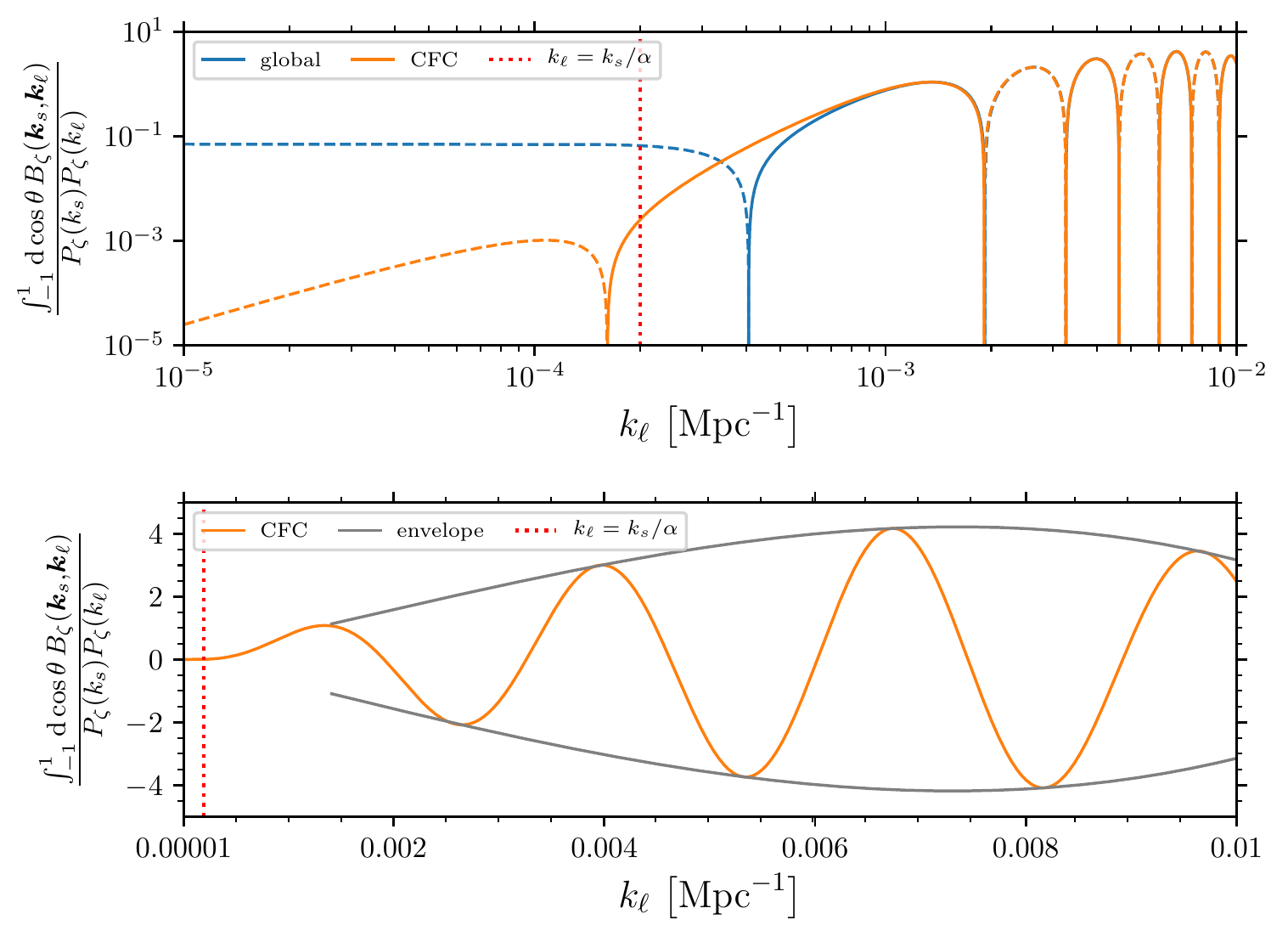} \\
\includegraphics[width=0.65\columnwidth]{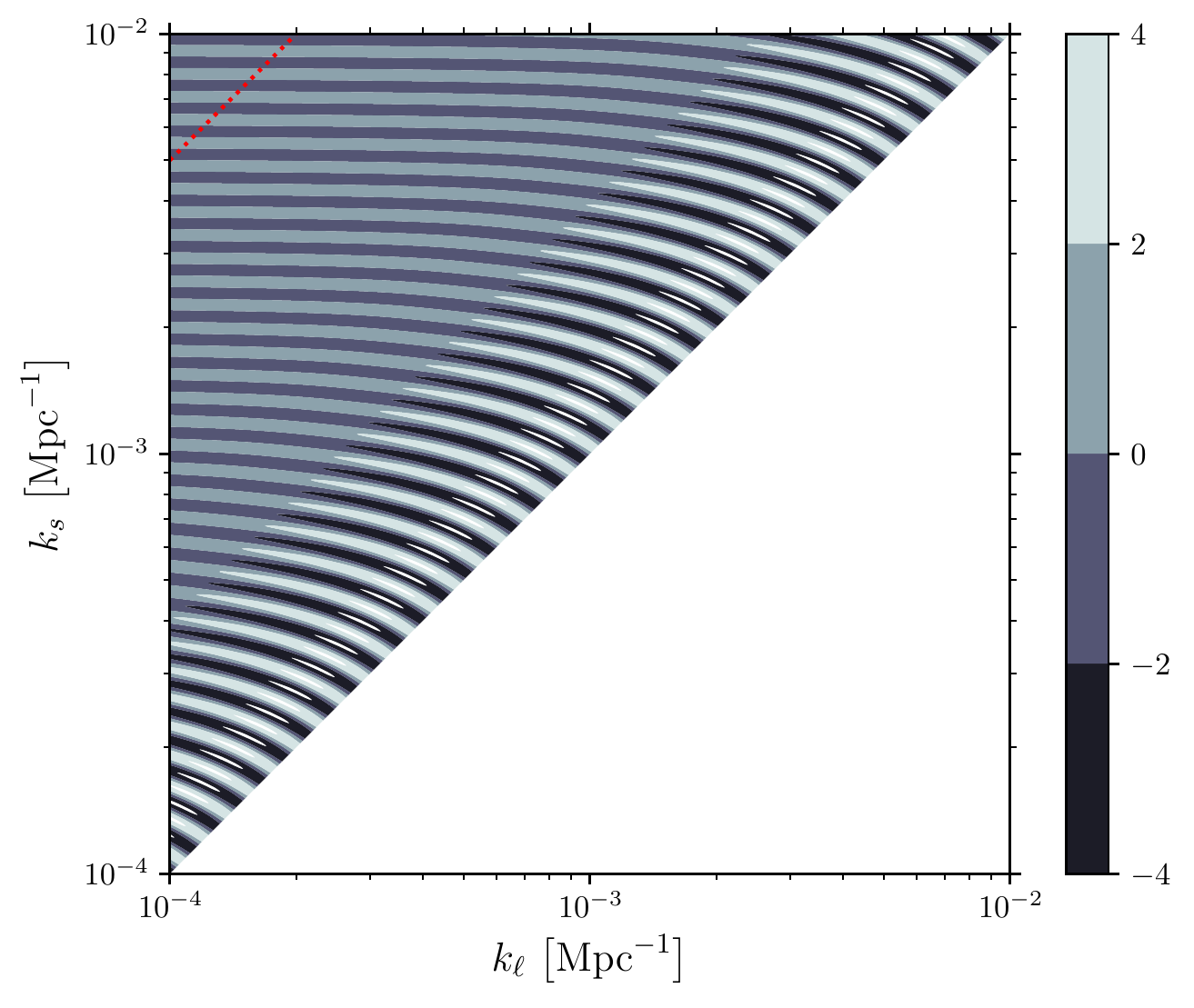}
\end{tabular}
\caption{The squeezed bispectrum of \eq{squeezed_CFC_bispectrum} for $\alpha=50$ and $k_\ast = \mpc{5e-2}$, 
integrated over $\cos\theta = \vers{k}_s\cdot\vers{k}_\ell$, for fixed $k_s$ (top panel) and as function of $(k_\ell,k_s)$ (bottom panel). 
As we can see from \eqsII{resonant_bispectrum}{enhancement}, the bispectrum is periodic under $k_\ast\to k_\ast\exp(2\pi n/\alpha)$, 
so we do not vary it in the plots. The two plots in the top panel have $k_s = \mpc{e-2}$ (dashed lines represent negative values): 
we see that in the intermediate squeezing regime the bispectrum in CFC (orange line) 
is similar to that in global coordinates (blue line). More precisely, it oscillates linearly with $k_\ell$ with a frequency proportional to $\alpha$, 
and has an envelope (shown as the grey line in the second plot of the top panel) different from a simple power law. 
For $k_\ell\lesssim k_s/\alpha$ (the red dotted lines show $k_\ell=k_s/\alpha$) we start to see a difference between the two bispectra, 
while for $k_\ell\ll k_s/\alpha$ the CFC one becomes $\propto k^2_\ell/k^2_s$ without oscillations in $k_\ell$, as shown in \eq{enhancement}. 
The contour plot in the bottom panel shows the full angle-averaged bispectrum for varying $k_\ell$ and $k_s$ 
(also here the red dotted line shows $k_\ell=k_s/\alpha$): 
we again see that in the ultra-squeezed regime $k_\ell\ll k_s/\alpha$ (upper-left corner of the plot) 
the response of the small-scale power spectrum to variations in $k_\ell$ goes quickly to zero, 
and the only oscillations are of the form $\cos(\alpha\log k_s)$. }
\label{fig:kernel_plot}
\end{figure}

We are now in position to discuss what happens in the case of resonant non-Gaussianity: 
to have an idea of how the basis of operators should be augmented in this case, 
we need to determine how the long mode modifies the small-scale power spectrum. 
Let us start by considering the ``ultra-squeezed'' limit $k_\ell/k_s\ll1/\alpha$, as opposed to the squeezed limit, $k_\ell/k_s\ll1$. 
In this limit, the leading effect of the long mode on the short-scale power spectrum is equivalent to that of equilateral non-Gaussianity, 
as we can see from the CFC bispectrum of \eq{enhancement}. That is, the small-scale power spectrum is modulated as in \eq{bias_exp-B} with $\Delta = 2$. 
However, things become more complicated as we take the long mode closer to $k_s/\alpha$. 
As we see from Fig.~\ref{fig:kernel_plot}, 
for $k_s\ll1/\alpha<k_\ell<k_s$ the response of the small-scale power spectrum to the long mode is made up of two contributions: 
a slowly-varying envelope and an oscillatory part. 
The oscillations are linear in $k_\ell$, and their frequency increases linearly with $\alpha$. 
Since the envelope is varying slowly with $k_\ell$, we can imagine that it will be described fully by an expansion of the form of \eq{bias_exp-D}. 
The same cannot be said about the oscillatory part: 
for it we also have a superposition of modulations of the form of \eq{bias_exp-D} (starting from $\Delta = 0$), 
but the coefficients $a_{0,2n}$ are now of order $\alpha^{2n}$: 
correspondingly, in the bias expansion we would have terms of the form 
\begin{equation}
\label{eq:bias_exp-O}
\delta_h\supset b_{\Psi^{(2n)}}\alpha^{2n}\Psi^{(2n)} \propto b_{\Psi^{(2n)}}\alpha^{2n}\partial^{2n}\zeta\,\,.
\end{equation}
Therefore, from \eq{bias_exp-E} we conclude that the bias expansion loses all predictive power 
due to the nontrivial behavior of the non-separable bispectrum in this model, 
since all terms beyond the ultra-squeezed limit become equally important.\footnote{Notice that, as in the usual case, 
there will be also effects coming from the transfer function. We expect them to still scale as powers of $(k/k_\text{eq})^2$, 
\ie without an enhancement at large $\alpha$, since they do not arise from the expansion of the bispectrum in the squeezed limit.}

However it is still possible to make some progress if we make the strong assumption that 
halo formation is an exactly local function of the initial density field smoothed on the single scale $R_\ast$. 
This happens, \emph{e.g.}, if we consider microscopic models of halo abundance: 
in this case one makes a specific ansatz about how precisely the Lagrangian halo number density depends on the statistics of the small-scale modes, 
and then assumes that the same relation holds when primordial non-Gaussianity is included. 
Since now there is no restriction to scales $k\ll R_\ast^{-1}$, this approach does not rely on expanding the primordial bispectrum in the squeezed limit. 
We can think of this approach as a resummation of all the beyond-squeezed-limit contributions discussed above. This will give rise to a 
scale-dependent correction $\Delta b_1(k)$ to the linear bias $b_1$, \ie $b_1\to b_1+\Delta b_1(k)$. 
For example, Ref.~\cite{Desjacques:2011mq} derived this correction by applying a conditional mass function approach 
(see \cite{Matarrese:1986et,Verde:2009hy} for related, previous approaches). 
The non-Gaussianity was taken into account by applying an Edgeworth expansion to the Gaussian PDF of the density field. 
Stopping the Edgeworth expansion at the bispectrum level, $\Delta b_1(k)$ is given by
\begin{equation}
\label{eq:bias_exp-P}
\Delta b_1(k) = \bigg[b_1\delta_\text{c} + \frac{\partial\log{\cal F}^{(3)}_\ast}{\partial\log\sigma_\ast}\bigg]\frac{2{\cal F}^{(3)}_\ast(k)}{\mathcal{M}_\ast(k)}\,\,.
\end{equation}
Here, the shape factor ${\cal F}^{(3)}_\ast$ is defined by (we drop the subscript $\ell$ on $k_3 = k_\ell$ from now on)
\begin{equation}
\label{eq:bias_exp-Q}
{\cal F}^{(3)}_\ast(k) = \frac{1}{4\sigma_\ast^2P_\phi(k)}\int\frac{\dif^3 k_s}{(2\pi)^3}\mathcal{M}_\ast(k_1)\mathcal{M}_\ast(k_2)B_\phi(k_1,k_2,k)\,\,,
\end{equation}
with $\vec{k}_1=\vec{k}_s-\vec{k}/2$, $\vec{k}_2={-\vec{k}_s}-\vec{k}/2$, and $\mathcal{M}_\ast(k)$ is equal to 
\begin{equation}
\label{eq:bias_exp-R}
\mathcal{M}_\ast(k)\equiv\mathcal{M}(k)W_\ast(k)\,\,,\qquad\mathcal{M}(k) = \frac{2}{3}\frac{k^2T(k)D_1(z)}{\Omega_\text{m}H^2_0}\,\,.
\end{equation}
Notice that in \eq{bias_exp-P} we can assume that $\mathcal{M}_\ast(k)\approx\mathcal{M}(k)$: 
this is a good approximation as long as we look at correlations on scales $k\ll 1/R_\ast$ (since we have $W_\ast(k)\approx 1$ there). 

Let us now investigate in more detail the meaning of \eqsII{bias_exp-P}{bias_exp-Q}. 
Using \eq{bias_exp-Q}, we can define the initial local variance-field of the small-scale density on the scale $R_\ast$ 
in the presence of long-wavelength potential perturbations $\phi(\vec{k})$ as 
\begin{equation}
\label{eq:bias_exp-S}
\hat{\sigma}^2_\ast(\vec{q})\equiv\sigma^2_\ast
\left[1 + 
\int\frac{\dif^3k}{(2\pi)^3}{\cal F}^{(3)}_\ast(k)\phi(\vec{k}) 
e^{i\vec{k}\cdot\vec{q}}\right]\,\,.
\end{equation}
The meaning of this local variance can be better understood if we look at \eq{bias_exp-B}. 
Indeed, $\hat{\sigma}^2_\ast(\vec{q})$ is the integral of the local power spectrum $P_\phi(\vec{k}_s|\vec{q})$ over $k_s$, weighted by $\mathcal{M}_\ast^2(k_s)$: 
that is, we can write the local power spectrum as
\begin{equation}
\label{eq:local_PS-A}
P_\phi(\vec{k}_s|\vec{q}) = \bigg[1+\int\frac{\mathrm{d}^3k}{(2\pi)^3}\mathcal{F}^{(3)}_\ast(k)\phi(\vec{k})e^{i\vec{k}\cdot\vec{q}}\bigg]P_\phi(k_s)\,\,.
\end{equation}
By comparing with \eq{bias_exp-B}, then, we see that the local power spectrum is now modulated by the operator 
$\int\frac{\mathrm{d}^3k}{(2\pi)^3}\mathcal{F}^{(3)}_\ast(k)\phi(\vec{k})e^{i\vec{k}\cdot\vec{q}}$, 
which must be then included in the bias expansion. 
We also notice that, using \eqsII{bias_exp-A}{bias_exp-D} at leading order in the squeezed limit, we have 
\begin{equation}
\label{eq:local_PS-B}
\begin{split}
{\cal F}^{(3)}_\ast(k) &= \frac{1}{4\sigma_\ast^2P_\phi(k)}\int\frac{\dif^3 k_s}{(2\pi)^3}\mathcal{M}_\ast(k_1)\mathcal{M}_\ast(k_2)B_\phi(k_1,k_2,k) \\
&\approx\frac{1}{4\sigma_\ast^2P_\phi(k)}\int\frac{\dif^3 k_s}{(2\pi)^3}\mathcal{M}_\ast^2(k_s)\bigg[4a_{0,0}\bigg(\frac{k}{k_s}\bigg)^\Delta P_\phi(k_s)P_\phi(k)\bigg] \\
&=\frac{a_{0,0}k^\Delta}{\sigma_\ast^2}\int\frac{\dif^3 k_s}{(2\pi)^3}k_s^{-\Delta}W^2_\ast(k_s)P_\text{m}(k_s) \\
&= a_{0,0}k^\Delta\frac{\sigma^2_{\ast,-\Delta/2}}{\sigma_\ast^2}\,\,,
\end{split}
\end{equation}
where $\Delta = 2$. That is, the additional operator $\int\frac{\mathrm{d}^3k}{(2\pi)^3}\mathcal{F}^{(3)}_\ast(k)\phi(\vec{k})e^{i\vec{k}\cdot\vec{q}}$ 
becomes degenerate with $\partial^2\phi\sim\delta_\text{m}$, \mbox{as expected.} 

Then, let us go back to a bispectrum of the form of \eq{bias_exp-A}, 
and let us focus on the leading order in the squeezed limit: 
then, the scale-dependent bias $\Delta b_1(k)$ would be given by (this is merely a rephrasing of $\delta_h\supset b_{\Psi^{(\Delta)}}\Psi^{(\Delta)}$)
\begin{equation}
\label{eq:bias_formula-A}
\Delta b_1(k)|_{\Delta}=a_{0,0}b_{\Psi^{(\Delta)}}k^\Delta\mathcal{M}^{-1}(k)\,\,,
\end{equation}
where $b_{\Psi^{(\Delta)}}$ is given by \eq{bias_exp-K} for a mass function of the form of \eq{bias_exp-G}. 
At higher orders in the squeezed limit, we would have
\begin{equation}
\label{eq:bias_formula-B}
\Delta b_1(k)|_{\Delta+2n}=a_{0,2n}b_{\Psi^{(\Delta+2n)}}k^{\Delta+2n}\mathcal{M}^{-1}(k)\,\,.
\end{equation}
Therefore we see that the (scale-dependent) resummation of all the bias coefficients $b_{\Psi^{(\Delta+2n)}}$ becomes
\begin{equation}
\label{eq:bias_formula-C}
\sum_{n=0}^{+\infty}a_{0,2n}b_{\Psi^{(\Delta+2n)}}k^{\Delta+2n} = \Delta b_1(k)\mathcal{M}(k) = 
\bigg[b_{\Psi^{(0)}} + 4\bigg(\frac{\partial\log\big(\sigma^2_\ast{\cal F}^{(3)}_\ast(k)\big)}{\partial\log\sigma_\ast^2}-1\bigg)\bigg]{\cal F}^{(3)}_\ast(k)\,\,,
\end{equation}
where ${\cal F}^{(3)}_\ast(k)$ is given by \eq{bias_exp-Q}, and in the last equality we used \eq{bias_exp-N}. 
Ref.~\cite{Desjacques:2011mq} showed that, 
for separable bispectrum shapes and for halos following a universal mass function, 
the leading squeezed-limit prediction of \eq{bias_exp-P} agrees with \eq{bias_exp-K}. 
Moreover, Ref.~\cite{Schmidt:2013nsa} showed that this agreement also holds at the next-to-leading order in the squeezed-limit expansion. 
Finally, a more detailed proof of \eq{bias_formula-C} and description of the resummation procedure are collected in Appendix \ref{app:resummation_proof}. 

After this discussion we can see that everything boils down to the integral of \eq{bias_exp-Q}. 
Since $\mathcal{M}_\ast(k_s)\propto k^2_sj_1(k_sR_\ast)$, this integral has support mainly for $k_s \gtrsim 1/R_\ast$, 
which following our assumptions is in the squeezed limit $k\ll k_s$. Therefore we can 
use the CFC bispectrum for $\phi$: using \eq{initial_conditions}, we see that it is related to that of $\zeta$ by
\begin{equation}
\label{eq:cfc_phi_bispectrum}
B_\phi^F(k_1,k_2,k)=\bigg({-\frac{3(1+w)}{5+3w}}\bigg)^3B_\zeta^F(k_1,k_2,k)\,\,,
\end{equation}
where we take $w=0$ since we are interested in short modes that re-enter the Hubble radius during matter dominance. 
Here the bispectrum $B_\zeta^F(k_1,k_2,k)$, whose angle average is shown in detail in Fig.~\ref{fig:kernel_plot}, is given by \eq{enhancement}. 

We now have all the ingredients to study the scale dependence of the halo bias in resonant non-Gaussianity. 
In the next section we compute $\Delta b_1$ for varying dimensionless frequency $\alpha$, 
halo mass $M$ from $10^{11}h^{-1}\,M_\odot$ to $10^{16}h^{-1}\,M_\odot$,\footnote{We 
compute the variation of the Jacobian in \eq{bias_exp-P} using the chain rule: 
\begin{equation*}
\frac{\partial\log{\cal F}^{(3)}_\ast}{\partial\log\sigma_\ast}=
\frac{\partial\log{\cal F}^{(3)}_\ast}{\partial\log M}\bigg[\frac{\partial\log\sigma_\ast}{\partial\log M}\bigg]^{-1}\,\,.
\end{equation*} }
and scale $k$ from $h\,10^{-4}\,\mathrm{Mpc}^{-1}$ to $h\,10^{-1}\,\mathrm{Mpc}^{-1}$. Notice that 
for the separate universe picture to work the long mode $k$ should be outside the (sound) horizon during radiation dominance. 
Therefore, we should restrict ourselves to $k\lesssim k_{\rm eq}$. 
The corrections to the bispectrum that come by taking $k$ shorter than $k_{\rm eq}$ are 
captured by the transfer function at second order in perturbations. 
For Gaussian initial conditions, these contributions to the bispectrum
give rise to $\fnlloc = {\cal O}(1)$ (see \cite{Fitzpatrick:2009ci} for a computation). 
We neglect these contributions in the following since they arise from different physics than inflation. 
The correction to the terms in the bispectrum proportional to $\fnlres$ from second-order evolution 
during and after radiation dominance, instead, will be very small and can be safely neglected. 
All plots will assume a flat $\Lambda$CDM cosmology with 
$\Omega_\mathrm{b} = 0.048$, 
$\Omega_\mathrm{c} = 0.258$, 
$H_0=\SI{67.7}{km.s^{-1}.Mpc^{-1}}$, 
$\log(\num{d10}\mathcal{A}_\mathrm{s})=3.067$, 
$n_\mathrm{s}=\num{0.967}$, 
$k_\ast=\mpc{0.05}$ \cite{Ade:2015xua}. 
We evaluate $\Delta b_1$ at $z=0$, 
using the Sheth-Tormen mass function \cite{Sheth:1999mn} to compute the Gaussian bias $b_1$ of \eq{bias_exp-M} 
and the Eisenstein-Hu transfer function \cite{Eisenstein:1997ik,Eisenstein:1997jh} to compute the linear matter power spectrum. 
Regarding the maximum value of $\alpha$, 
we notice that a practical constraint arises from the fact that, in the limit $\alpha\gg 1$, 
the bispectrum oscillates rapidly and the numerical integration to obtain $\Delta b_1$ becomes difficult. 
For this reason, we will go up to $\alpha = 20$ in our numerical analysis 
(and, consistently with Fig.~\ref{fig:fig_40_planck}, we take $\fnlres = \num{3d-4}\times\alpha^{\num{2.63}}$): 
as we will see in the next section, this is enough to estimate the scaling of $\Delta b_1$ with $k$ and $\alpha$.

\subsection{Results}
\label{sec:resonant_results}

\begin{figure}[H]
\myfloatalign
\centering
\begin{tabular}{c}
\includegraphics[width=0.65\columnwidth]{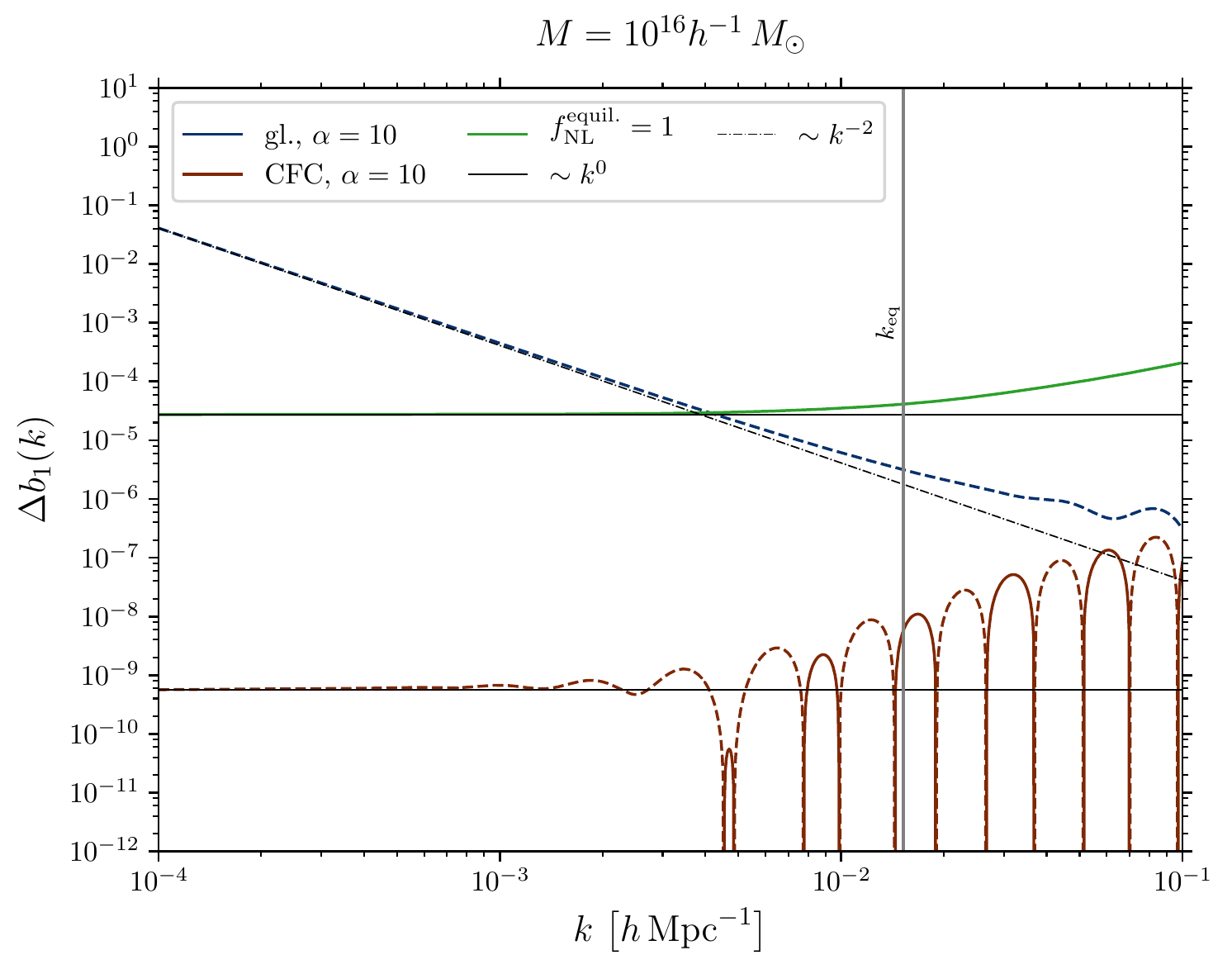} \\
\includegraphics[width=0.65\columnwidth]{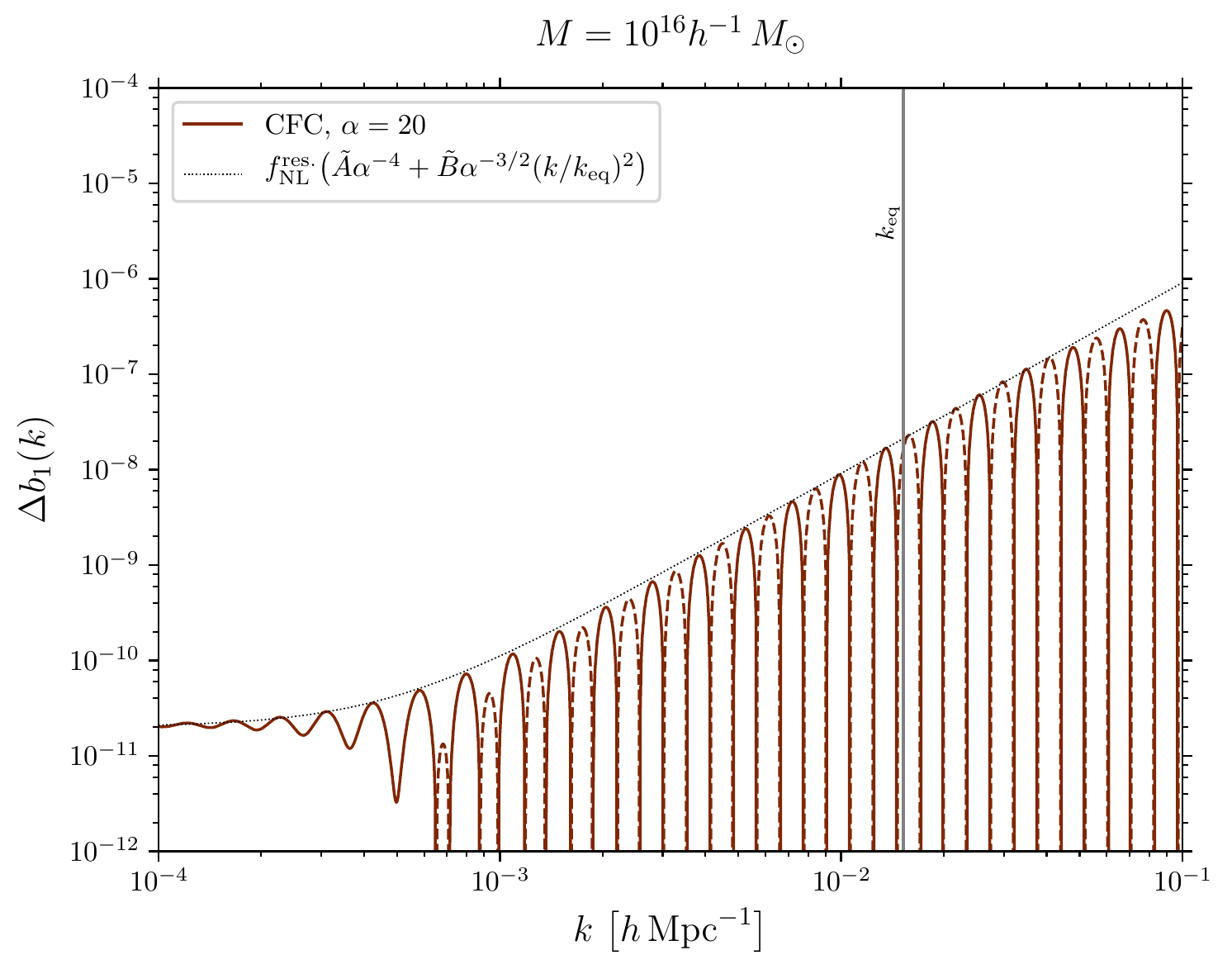}
\end{tabular}
\caption{Top panel: non-Gaussian halo bias in global coordinates (blue line) and CFC (red line) for $\alpha=10$ 
as a function of $k$ at $M = \num{e16}{h^{-1}\,M_\odot}$, 
together with that for equilateral non-Gaussianity (green line) for $f_\text{NL}^\text{equil}=1$ 
\cite{Schmidt:2010gw,Desjacques:2011mq,Desjacques:2016bnm}. 
The latter can be converted into the prediction of single-field slow-roll models, 
for which $f_\text{NL}^\text{equil}\sim\num{d-2}$ \cite{Cabass:2016cgp}, 
or that of $P(\phi,X)$ theories with non-canonical speed of sound $c^2_\text{s}\neq 1$, 
that have $f_\text{NL}^\text{equil}\sim(1-c^2_\text{s})/c^2_\text{s}$ \cite{Creminelli:2013cga,Cabass:2016cgp}. 
Dashed lines indicate negative values of $\Delta b_1$. 
Bottom panel: $\Delta b_1(k)$ in CFC for $\alpha=20$ and $M = \num{e16}{h^{-1}\,M_\odot}$, together with the fit of \eq{fit} (black dotted line). 
The vertical line denotes the point at which the transfer function $T(k)$, entering in the definition of $\Delta b_1$ through $\mathcal{M}(k)$ in \eq{bias_exp-P}, 
starts to affect its scale dependence: 
from the bottom panel, we see that for $k/k_\text{eq}\approx 10$ this causes the scaling of the non-Gaussian bias to slightly deviate from $\sim k^2$. }
\label{fig:resonant-varying_k}
\end{figure}

\noindent We start from studying the dependence of $\Delta b_1$ on $k$ at fixed halo mass $M = \num{e16}{h^{-1}\,M_\odot}$. 
The top panel of Fig.~\ref{fig:resonant-varying_k} shows that, 
while for very long $k$ the CFC transformation makes the bias scale-independent 
(as opposed to $\Delta b_1(k)\sim k^{-2}$, which is the scaling obtained using the bispectrum in global coordinates), 
it oscillates with $k$ on smaller scales. As $\alpha$ becomes larger, these oscillations begin at a longer $k$. 
We see that these oscillations eventually become logarithmic in $k$, and have a frequency that increases linearly with $\alpha$. 
We can also fit for the envelope of the oscillations: as we see in the bottom panel of Fig.~\ref{fig:resonant-varying_k}, 
the envelope is very well approximated by
\begin{equation}
\label{eq:fit}
\abs{\Delta b_1(k)}\leq\fnlres\bigg[\frac{\tilde{A}}{\alpha^4}+\frac{\tilde{B}}{\alpha^{3/2}}\bigg(\frac{k}{k_\text{eq}}\bigg)^2\bigg]\,\,,
\end{equation}
where $\tilde{A}$ and $\tilde{B}$ depend very weakly on $\alpha$ 
(the details of their dependence on $\alpha$, 
together with a plot of $\Delta b_1$ for $\alpha=50$, are shown in Appendix \ref{app:appendix_B}). 
In practice, the constant contribution $\propto\tilde{A}/\alpha^4$ to $\Delta b_1(k)$ is absorbed in the Gaussian large-scale bias, and is thus unobservable. 
Hence, the observable scale-dependent bias satisfies 
\begin{equation}
\label{eq:fit_abs}
\abs{\Delta b_1(k)}_\text{obs}\leq\frac{\fnlres{\tilde{B}}}{\alpha^{3/2}}\bigg(\frac{k}{k_\text{eq}}\bigg)^2\,\,,
\end{equation}
where ${\tilde{B}(\alpha)}$ oscillates around $\num{2.7d-6}$.

\begin{figure}[H]
\myfloatalign
\centering
\begin{tabular}{c}
\includegraphics[width=0.65\columnwidth]{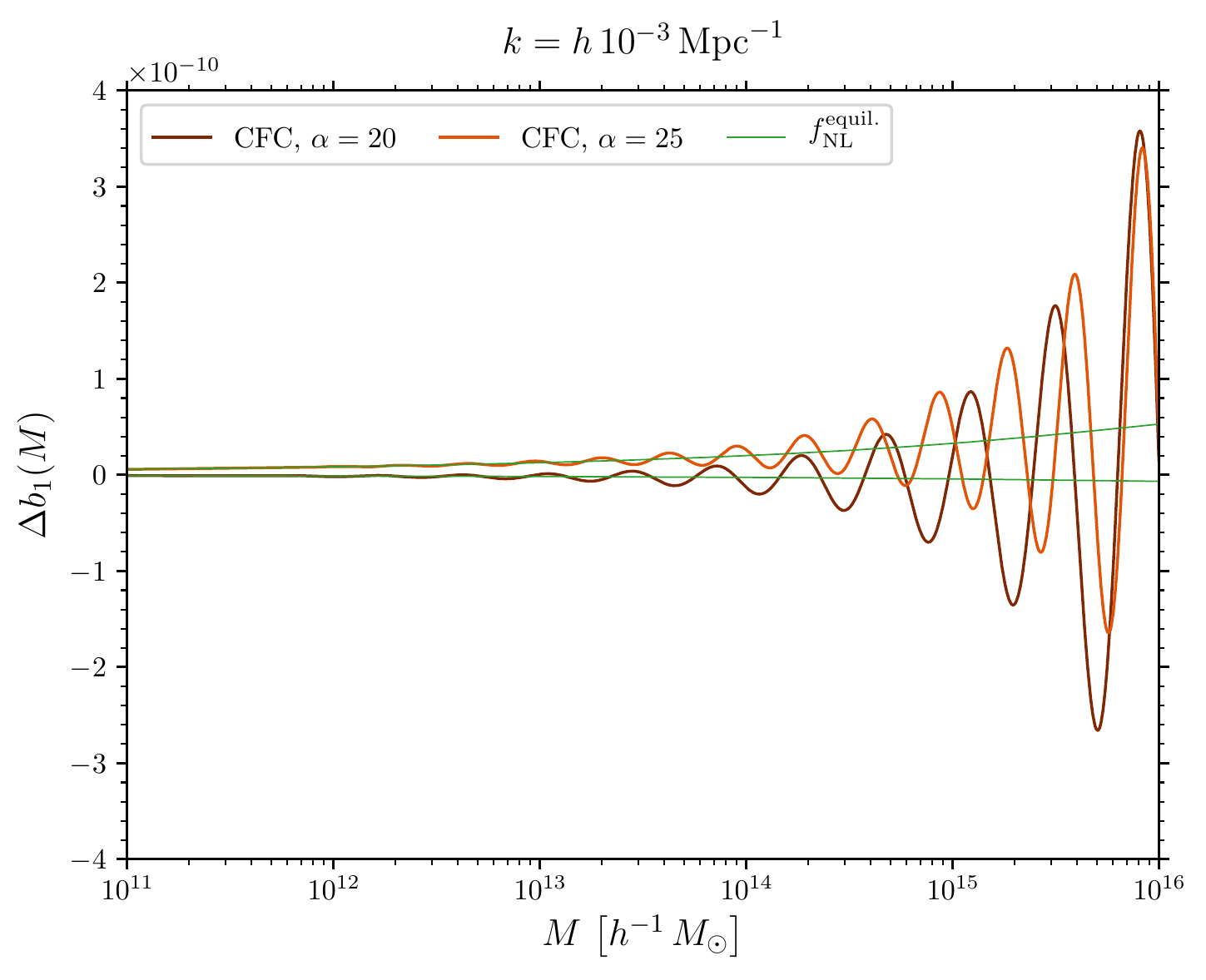} \\
\includegraphics[width=0.65\columnwidth]{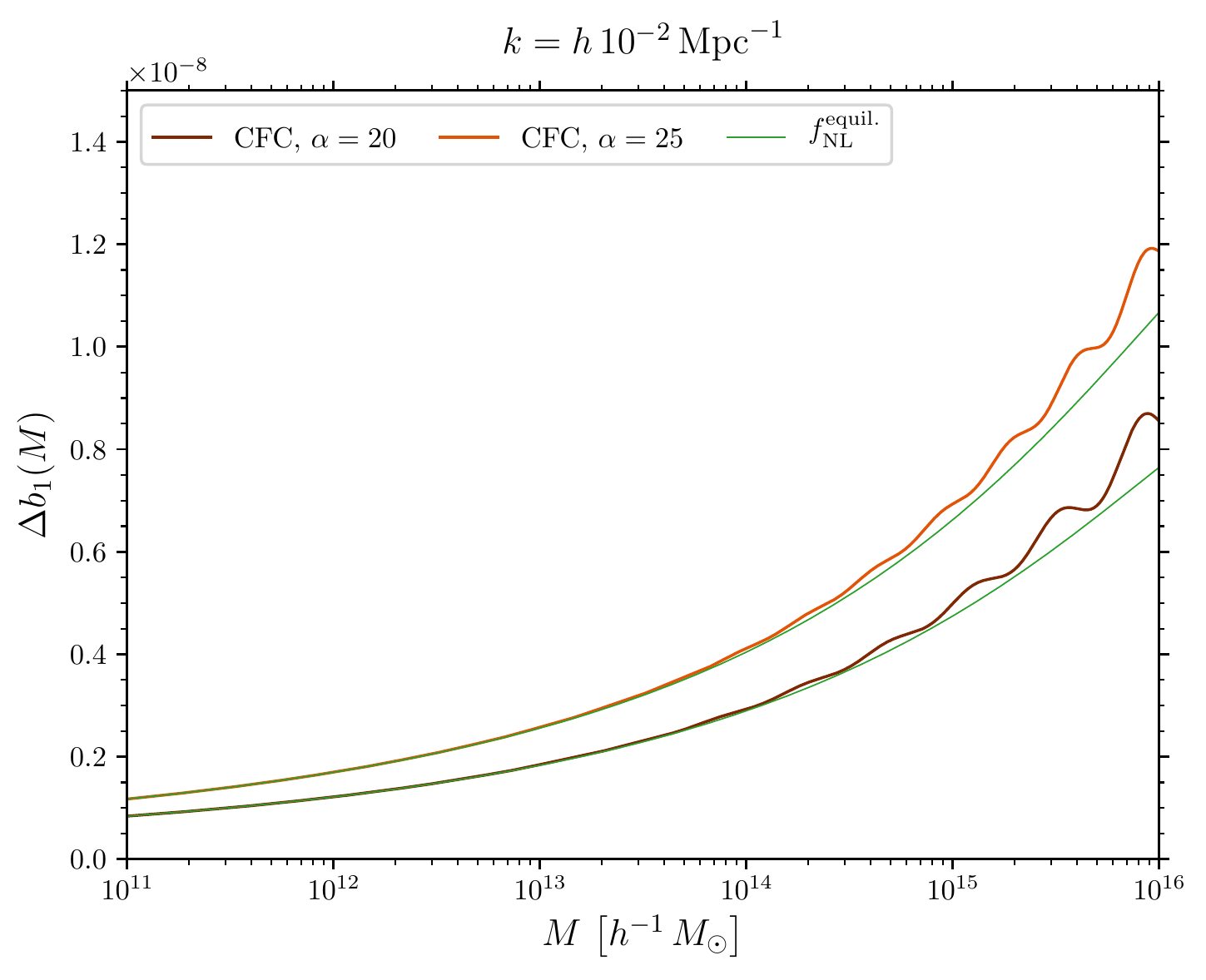}
\end{tabular}
\caption{Non-Gaussian halo bias correction in CFC as a function of halo mass for $k= h\,\mpc{e-3}$ (top panel) and $k=h\,\mpc{e-2}$ (bottom panel): 
we see that taking $k$ near to $k_\text{eq}\approx\mpc{e-2}$ substantially diminishes the amplitude of the oscillations. 
However, we see that in both cases the envelope of the oscillations is an increasing function of the halo mass: 
comparing with the green lines, we see that the bias increases with $M$ faster than equilateral non-Gaussianity. 
Notice that in these plots we normalize $\fnleq$ to match $\Delta b_1$ at $M = \num{e11}{h^{-1}\,M_\odot}$, 
in order to better compare the dependence on the halo mass. 
A more precise discussion is presented in Section \ref{sec:observations}.}
\label{fig:resonant-varying_M}
\end{figure}

Then, we revisit the dependence of the non-Gaussian bias on the halo mass: 
in \cite{CyrRacine:2011rx} it was shown that on large scales the resonant non-Gaussianity bispectrum of \eq{resonant_bispectrum} 
predicts a strong oscillatory dependence on $M$ of the halo bias, over a broad range of masses. 
In Fig.~\ref{fig:resonant-varying_M} we see that such behavior was mainly due to the unphysical contribution from the consistency relation: 
if the CFC bispectrum of \eq{squeezed_CFC_bispectrum} is used to compute $\Delta b_1$, 
the oscillations with $M$ are suppressed. 
Increasing $\alpha$, however, 
causes the modulation with $M$ to show up over the whole range $\num{e11}{h^{-1}\,M_\odot}\leq M\leq\num{e16}{h^{-1}\,M_\odot}$. 
For $k\approx k_\text{eq}$, we see that the amplitude of the oscillations with $M$ becomes much smaller. 
Nevertheless, we confirmed numerically that the contribution to $\Delta b_1$ 
coming from the effect of the long mode on the Jacobian $\abs{\dif\log\sigma_\ast/\dif\log M}$, 
\ie the second term in the square bracket of \eq{bias_exp-P}, 
is always dominant over the one coming from the rescaling of the threshold for spherical collapse 
(\ie the term $b_1\delta_\text{c}$ in \eq{bias_exp-P}).

There is an additional parameter in the resonant bispectrum of \eq{resonant_bispectrum} that we can vary, \ie 
the pivot scale $k_\ast$. 
As it is clear from \eqsII{resonant_bispectrum}{enhancement}, 
there is periodicity in the non-Gaussian bias under $k_\ast\to k_\ast\exp(2\pi n/\alpha)$: 
More precisely, $\Delta b_1$ changes sign for $k_\ast\to k_\ast\exp(\pi n/\alpha)$. 
We have checked that the numerical result reproduces the expected periodicity. 
Besides, we have checked that changing the pivot scale within the range $\big[k_\ast,k_\ast\exp(\pi/(2\alpha))\big]$ 
does not lead to quantitative differences in the overall amplitude of $\Delta b_1$ 
or in its dependence on $k$. We conclude by noting that varying $k_\ast$ corresponds to 
considering an additional phase $\varphi$ in \eq{resonant_potential}, \ie
\begin{equation}
\label{eq:potential_with_phase}
V(\phi) = V_0(\phi) + \Lambda^4\cos\bigg(\frac{\phi}{f}+\varphi\bigg)\,\,.
\end{equation}
Indeed, the calculation of \cite{Flauger:2010ja}, that has been carried out for $\varphi = 0$, 
shows that the trigonometric functions appearing in the bispectrum have as argument ${\phi_{k_t}}/{f}$, 
where $k_t\equiv\sum_{i=1}^3 k_i$ and 
$\phi_k = \phi_\ast - \sqrt{2\eps_\ast}\log(k/k_\ast)$ is the value of the field when the mode of momentum $k$ crosses the Hubble radius. 
Up to slow-roll corrections, then, 
we can map those results to the generic potential of \eq{potential_with_phase} by replacing ${k_\ast}$ with ${k_\ast\exp{(\varphi/\sqrt{2\eps_\ast})}}$.

\subsection{Observational prospects}
\label{sec:observations}

\noindent The most relevant result of the previous section is the dependence of $\Delta b_1$ on $k$: 
while in global coordinates the scaling is indistinguishable from that of local non-Gaussianity, 
once the unphysical contribution from the consistency relation is subtracted, the bias shows a strong oscillatory behavior, 
with frequency of oscillations $\propto\alpha$ (as seen in the bottom panels of Figs.~\ref{fig:resonant-varying_k}, \ref{fig:additional_plots}). 
This behavior is radically different from all the types of primordial non-Gaussianity considered in the literature, 
and it can potentially make searches of resonant non-Gaussianity in the halo bias a powerful complement to the current bounds from the CMB: 
indeed, CMB data have a strong constraining power mainly for low values of $\alpha$. 

We now carry out an approximate, simple forecast of the detectability of the non-Gaussian bias. 
First, note that the $1\sigma$ error on $\fnlres$, for a fiducial value of $\fnlres=0$ and keeping all other parameters including $b_1$ fixed, 
is given by\footnote{At linear order, and neglecting effects such as redshift-space distortions and shot noise which do not depend on $\fnlres$, 
we can write $P_h(k) = [b_1+\Delta b_1(k)]^2P_m(k)$. Then, the derivative with respect to $\fnlres$ at $\fnlres=0$ becomes
\begin{equation*}
\frac{\partial P_{h}(k)}{\partial\fnlres} = \frac{\partial\Delta b_1(k)}{\partial\fnlres}\frac{\partial P_h(k)}{\partial\Delta b_1(k)}\approx 
\frac{\partial\Delta b_1(k)}{\partial\fnlres}2b_1P_m(k)\approx\frac{\partial\Delta b_1(k)}{\partial\fnlres}\frac{2P_h(k)}{b_1}\,\,.
\end{equation*}
}
\begin{equation}
\sigma^2(\fnlres) = \sum_{\vec{k}}\bigg(\frac{\partial P_{h}(k)}{\partial\fnlres}\bigg)^{-2} \sigma^2[P_h(k)]
\approx\sum_{\vec{k}}\bigg(\frac{\partial \Delta b_1(k)}{\partial\fnlres}\bigg)^{-2} \frac{\sigma^2[P_h(k)]}{[2P_h(k)/b_1]^2}\,\,,
\end{equation}
where the sum runs over all wavenumbers observable in the survey (up to some $k_{\rm max}$), 
$P_h(k)$ is the fiducial halo power spectrum, and $\sigma^2[P_h(k)]$ is its variance which is due to cosmic variance and shot noise. 
It is then clear that the detection threshold for $\fnlres$ from the scale-dependent bias $\Delta b_1(k)$ can be roughly estimated using its envelope in \eqsII{fit}{fit_abs}. 
That is, the detection significance is not increased by the presence of the oscillations.
This allows us to carry out a quick forecast on the detectability of the non-Gaussian bias by matching its scale-dependence to that of equilateral non-Gaussianity.

Let us be more precise: in equilateral non-Gaussianity, while on very large scales the bias is scale-independent, 
as we take $k$ close to $k_\text{eq}$ some scale dependence arises due to the transfer function. 
Consider the bispectrum of equilateral non-Gaussianity, \ie
\begin{equation}
\label{eq:equil_bispectrum}
\begin{split}
&B_\zeta^{\text{equil}}(k_1,k_2,k_3) = 
6\fnleq\big[{-P_\zeta(k_1)P_\zeta(k_2)}-\text{$2$ perms.} - 2\big(P_\zeta(k_1)P_\zeta(k_2)P_\zeta(k_3)\big)^{2/3} \\
&\hphantom{B_\zeta^{\text{equil}}(k_1,k_2,k_3) = 6\fnleq\big[} + P_\zeta(k_1)^{1/3}P_\zeta(k_2)^{2/3}P_\zeta(k_3) + \text{$5$ perms.}\big]\,\,,
\end{split}
\end{equation}
and let us focus on the leading order in the squeezed limit, \ie \eq{bias_exp-D} for $n=0$. We see that $\Delta=2$ and $a_{0,0}=-10\fnleq/3$. 
We can then use \eqsIII{bias_exp-J}{bias_exp-K}{bias_formula-A} to compute $\Delta b_1(k)|_{\Delta=2}$, 
which is plotted as a green line in the top panel of Fig.~\ref{fig:resonant-varying_k} and in Fig.~\ref{fig:resonant-varying_M}. 
Expanding the inverse transfer function in powers of $k/k_\text{eq}$, the scale dependence takes the form
\begin{equation}
\label{eq:equil_bias-A}
\Delta b_1(k)|_{\Delta=2}= {-\frac{{5\fnleq}\,\Omega_\text{m}H^2_0\,b_{\Psi^{(2)}}}{{T(k)D_1(z)}}}
\equiv\fnleq \bigg[A_\text{equil} + B_\text{equil}\bigg(\frac{k}{k_\text{eq}}\bigg)^2+\dots\bigg]\,\,.
\end{equation}
While the constant part is completely degenerate with $b_1$, 
we see that the scale dependence due to the transfer function is suppressed by $k_\text{eq}$, and not by $R_\ast$ 
as it happens for higher-derivative biases (\ie $\delta_h\supset b_{\partial^{2m}\delta_\text{m}}{\partial^{2m}\delta_\text{m}}$) 
or for beyond-squeezed-limit contributions from \eq{bias_exp-D}. 
Therefore the degeneracy between these different contributions can be broken if the scale dependence at $k\gtrsim k_\text{eq}$ 
can be measured with sufficient precision, since the different scales $R_\ast$ and $k_\text{eq}$ can be disentangled \cite{Assassi:2015fma,Desjacques:2016bnm}. 
This has been used in \cite{Gleyzes:2016tdh} to show that, with optimistic assumptions, 
future surveys could obtain $\sigma(\fnleq) = \mathcal{O}(\num{d2})$ from constraints on 
the galaxy power spectrum. 

Our goal, then, is to translate our fit of \eqsII{fit}{fit_abs} to an ``equivalent'' $\hat{f}_{\rm NL}^{\rm equil}$ 
by matching the scale-dependent parts of \eqsII{fit}{equil_bias-A}. That is, we define $\hat{f}_{\rm NL}^{\rm equil}$ as 
\begin{equation}
\label{eq:equil_bias-B}
\hat{f}_{\rm NL}^{\rm equil} = \frac{\tilde{B}}{B_\text{equil}}\frac{\fnlres}{\alpha^{3/2}}\,\,.
\end{equation}
We do this matching at $M = \num{e16}{h^{-1}\,M_\odot}$ and $z=0$.\footnote{The choice of redshift 
does not matter, since \eqsII{bias_exp-P}{bias_formula-A} scale with $z$ in the same way.} 
This is a conservative choice, since from Fig.~\ref{fig:resonant-varying_M} we see that 
the bias in resonant non-Gaussianity increases with mass faster than in equilateral non-Gaussianity. 
Doing the matching at $M < \num{e16}{h^{-1}\,M_\odot}$ would then result in a smaller $\hat{f}_{\rm NL}^{\rm equil}$, since the 
ratio $\tilde{B}/B_\text{equil}$ will be smaller. 
Recalling that $\tilde{B}\approx\num{2.7d-6}$ for $M = \num{e16}{h^{-1}\,M_\odot}$, 
and using the result of Fig.~\ref{fig:fig_40_planck} for $\fnlres$, we see that 
\begin{equation}
\label{eq:equil_bias-C}
\hat{f}_{\rm NL}^{\rm equil}\approx\num{1.7d-4}\alpha^{1.13}\,\,.
\end{equation}
The equivalent $\fnleq$ becomes of order $\num{0.5}$ for $\alpha = e^7\approx\num{1.1d3}$ 
(\ie the maximum value considered in the \emph{Planck} analysis): 
from this, we can conclude that measurements of the galaxy power spectrum from 
upcoming surveys will be unable to improve the CMB bounds on resonant non-Gaussianity, 
since future surveys can only achieve $\sigma(\fnleq) = \mathcal{O}(\num{d2})$ \cite{Gleyzes:2016tdh}.

Finally, let us comment on the relevance of the oscillations. As we discussed in Section \ref{sec:resonant_results}, 
at small scales $k\gtrsim k_\text{eq}$ the bias oscillates logarithmically with scale, \ie $\Delta b_1\sim\cos(\alpha\log k)$. 
If a scale-dependent bias due to resonant non-Gaussianity were to be detected, 
these oscillations would provide a very precise constraint on $\alpha$, if multiple oscillations can be measured. 
Notice, however, that the experimental resolution in $k$ is given by $\Delta k = 2\pi/L_\text{survey}$, 
where $L_\text{survey}$ is the typical comoving size of the survey considered: 
therefore, for logarithmic oscillations the relevant resolution is $\Delta k/k = 2\pi/(kL_\text{survey})$, 
which saturates to $(\Delta k/k)|_{\rm max} = 2\pi/(k_{\rm max}L_\text{survey})$, 
where $k_{\rm max}$ is the maximum wavenumber accessible by the survey. 
If $\alpha\gg\pi/(\Delta k/k)|_{\rm max}$, it is not possible to resolve the oscillations.

\section{Cosmological colliders}
\label{sec:QSF_intro}

\noindent We now move to the study of cosmological colliders models. 
We start by deriving the CFC bispectrum, and then we compute the bias $\Delta b_1$ following the approach of Section \ref{sec:bias_resonant_NG}.

\subsection{CFC bispectrum and bias expansion}
\label{sec:QSF_CFC}

\noindent In \cite{Lee:2016vti}, the coupling of massive spinning fields to the inflaton has been studied 
using the framework of the Effective Field Theory of Inflation. 
Consider for example a totally symmetric, traceless and massive spin-$s$ field $\sigma_{\mu_1\dots\mu_s}$. 
After the St\"{u}ckelberg trick, operators like $(\delta g^{00})^n(\sigma^{0\dots0})^m$ 
couple the Goldstone boson of broken time diffeomorphisms $\pi$ to $\sigma_{\mu_1\dots\mu_s}$. 
In order for the massive field to contribute to the scalar three-point function at tree level, 
a quadratic mixing $\delta g^{00}\sigma^{0\dots0}$ is needed. 
This affects the linearized equation of motion for the massive field: 
for masses much larger than Hubble, where we can integrate it out, 
we get additional local self-interactions of the Goldstone $\pi$, suppressed by powers of $\Box/m^2$. 
While for $s\neq 0$ they are not completely degenerate with the standard self-interactions of $\pi$ 
(since they still carry information on the spin of the integrated-out particle through the derivative structure), 
the information about the mass of the particle is lost since the mass dependence becomes degenerate with the coupling constant. 
Moreover, since they are local interactions, they give rise to bispectra which are analytic in Fourier space. 
Integrating out the massive field does not capture, however, all the effects: 
indeed, particles can be spontaneously created in a time-evolving background, 
an effect which cannot be represented by adding a local vertex to the effective Lagrangian \cite{Arkani-Hamed:2015bza}. 
The resulting bispectra, then, have a non-analytic dependence on momenta. 
In the following, we focus on this non-analytic shape: 
we refer to \cite{MoradinezhadDizgah:2018ssw} (see \eg their Appendix C) for a forecast that includes both local and non-local effects. 

For a given spin $s$ and mass $m$, 
the non-analytical scalings in the squeezed limit take the form \cite{Arkani-Hamed:2015bza,Lee:2016vti}
\begin{equation}
\label{eq:QSF_bispectrum}
\begin{split}
&\lim_{k_3\ll k_1\sim k_2}B_\zeta(k_1,k_2,k_3)\supset 
\frac{C_sf^{(s)}}{(2\pi^2)^2\Delta_\zeta}\bigg[\bigg(\frac{k_3}{k_1}\bigg)^{\frac{3}{2}}
\mathcal{L}_s(\vers{k}_1\cdot\vers{k}_3)\cos\bigg(\mu_s\log\frac{k_3}{k_1}+\phi_s\bigg)\bigg]P_\zeta(k_1)P_\zeta(k_3) \\
&\hphantom{\lim_{k_3\ll k_1\sim k_2}B_\zeta(k_1,k_2,k_3)\supset } + \vec{k}_1\to\vec{k}_2\,\,,
\end{split}
\end{equation}
where $\Delta_\zeta=\sqrt{\mathcal{A}_\text{s}}$. 
We defined $\mu_s = \sqrt{(m/H)^2 - (s-1/2)^2} \geq 0$ for $s\neq 0$: 
a minimally coupled scalar $\sigma^0$ has $\mu_0 = \sqrt{(m/H)^2 - (3/2)^2}$. 
The phase $\phi_s$ and the factor $f^{(s)}$ 
(which gives the overall Boltzmann suppression) 
are uniquely fixed in terms of $s$, $\mu_s$, and the speed of sound $c_\pi$ of the Goldstone boson of broken time diffeomorphisms 
(we follow the notation of \cite{Lee:2016vti} for the speed of sound). 
$C_s$, instead, depends on the strength of the quadratic mixing between $\pi$ and the spinning fields, 
and what kind of cubic vertex one is considering in the interaction Hamiltonian \cite{Lee:2016vti}. 

From \eq{QSF_bispectrum} we see that the spin gives an effect non-degenerate with the mass only through the overall Legendre polynomials: 
since the bias is sensitive only to the angle-averaged squeezed bispectrum we focus on zero spin 
(observables that are sensitive to the angular dependence of the primordial squeezed bispectrum are, for example, 
galaxy shapes \cite{Schmidt:2015xka,Chisari:2016xki}, galaxy alignments \cite{Kogai:2018nse} and, 
in general, the galaxy bispectrum \cite{Assassi:2015fma,MoradinezhadDizgah:2018ssw}).
The functions $f^{(s)}$ and $\phi_s$, for $s=0$, take the form \cite{MoradinezhadDizgah:2017szk}
\begin{subequations}
\label{eq:A_s_phi_s}
\begin{align}
&f^{(0)} = \frac{\pi^3}{2}\abs[\bigg]{\frac{1+i\sinh(\pi\mu)}{\cosh(\pi\mu)}\frac{\Gamma(-i\mu)}{\Gamma(1/2-i\mu)}}\times
\begin{cases}
\text{${-\dfrac{\pi^{3/2}}{8}}(1+4\mu^2)\sech(\pi\mu)$ for $c_\pi=1\,\,,$} \\[1em]
\text{${-\Gamma\bigg(\dfrac{3}{4}-\dfrac{i\mu}{2}\bigg)\Gamma\bigg(\dfrac{3}{4}+\dfrac{i\mu}{2}\bigg)}$ for $c_\pi\ll 1\,\,,$}
\end{cases}
\label{eq:A_s_phi_s-1} \\
&\phi_0 = \arg\bigg(\frac{\pi^3}{2}{\frac{1+i\sinh(\pi\mu)}{\cosh(\pi\mu)}\frac{\Gamma(-i\mu)}{\Gamma(1/2-i\mu)}}\bigg) 
- \mu\log(4c_\pi)\,\,, \label{eq:A_s_phi_s-2}
\end{align}
\end{subequations}
where we called $\mu_0\equiv\mu$ (not to be confused with the notation for $\vers{k}_s\cdot\vers{k}_\ell$ usually employed in the literature) 
and we allowed for the possibility of a small speed of sound $c_\pi\ll 1$, investigated in \cite{Lee:2016vti}. 
The amplitude $C_0$, instead, will be a function (at leading order in derivatives) of the coefficient of $\delta g^{00}\sigma$ 
and of the coefficients of the operators $(\delta g^{00})^2\sigma$, $\delta g^{00}\sigma^2$, and $\sigma^3$ 
(depending on what cubic vertex in the interaction Hamiltonian one is considering).

Given the squeezed bispectrum of \eq{QSF_bispectrum}, 
we can ask what happens once we switch to CFC and 
how we can relate it to the initial conditions for the Newtonian potentials at Hubble re-entry 
(\emph{i.e.}, what are the differences, if any, with the results of Section \ref{sec:bias_and_curvature}). 
As long as the mass of the additional fields is different from zero, 
the consistency relation is satisfied by the global bispectrum and 
the curvature perturbation $\zeta$ becomes a constant on super-Hubble scales 
(this has been shown in many works: 
see \cite{Chen:2009zp,Chen:2012ge,Lee:2016vti} for details). 
In this case, then, the results of Appendix \ref{app:appendix_A} will apply. 
However, we still need to compute the CFC squeezed bispectrum in order to make use of \eq{app_A-I}. 

The change from global coordinates to CFC involves both 
spatial and time coordinates, 
since the conformal time of comoving observers is not the same as that of the CFC ones. 
For an isotropic long mode, the spatial coordinate change just gets rid of the consistency condition in the global bispectrum, 
as discussed in Section \ref{sec:bias_and_curvature}. 
However, even if slow-roll suppressed contributions are neglected, 
the change in the time coordinate leads to a contribution to the CFC bispectrum proportional 
to the time derivative of the short-scale power spectrum. 
In the single-field case this contribution is trivially zero at late times. 
In the cosmological collider case the operator $\delta g^{00}\sigma$ couples $\zeta = {-H\pi}$ and $\sigma$ at 
quadratic level, modifying the time-dependence of the short-scale power spectrum with respect to the single-field case: 
it is then important to check whether this modification leads to additional terms appearing in the CFC bispectrum, 
or if it also vanishes at late times. 
Here we show that, as long as $\sigma$ has a non-zero mass, this contribution vanishes. 

The modification to the short-scale curvature perturbation 
coming from the change of the time coordinate $\eta = \eta_F + \xi^0_\ell(\eta_F,\vec{q})$, 
where $\xi^0_\ell(\eta_F,\vec{q})$ starts at first order in the long modes, 
takes the form
\begin{equation}
\label{eq:app_D_intro-A}
\zeta^F_s(\eta,\vec{q}) = \zeta_s(\eta,\vec{q}) - 
\frac{\partial_i\xi^0_\ell(\eta,\vec{q})\partial_i\zeta_s(\eta,\vec{q})}{3\cH} + \xi^0_\ell(\eta,\vec{q})\partial_0\zeta_s(\eta,\vec{q})\,\,,
\end{equation}
where we have neglected slow-roll-suppressed terms and we have dropped the $F$ subscript on the time coordinate. 
Let us now focus for simplicity only on the last term on the right-hand side of \eq{app_D_intro-A}: 
neglecting the second term will not change our conclusions, 
since its time-dependence is qualitatively the same as the last term. 
Without loss of generality, we can expand the long-wavelength $\xi^0_\ell(\eta,\vec{q})$ in a Taylor series in $\vec{q}$ around $\vec{0}$: 
it then takes the form 
(we drop the subscripts $\ell$ and $s$ to simplify the notation) 
\begin{equation}
\label{eq:app_D_intro-B}
\xi^0(\eta,\vec{q}) = \xi^0(\eta,\vec{0}) + A^0_i(\eta)\,q^i + B^0_{ij}(\eta)\,q^iq^j + C^0_{kij}(\eta)\,q^i q^j q^k\,\,.
\end{equation}
The uniform shift $\xi^0$ is constructed from the local Hubble rate $H_F = \nabla_\mu U^\mu/3$, where $U^\mu$ is the CFC observer, 
and the large-scale velocity divergence $\partial_i V^i$ (see \eg Appendix A of \cite{Cabass:2016cgp}). 
The coefficients $A^0_i$, $B^0_{ij}$ and $C^0_{kij}$ 
are also constructed from the long-wavelength metric evaluated at $\vec{q}=\vec{0}$: 
for example, we have that $A_i^0$ is equal to $V_i - \partial_i\zeta/\cH$ up to slow-roll suppressed terms, 
while $B^0_{ij}$ contains $\partial_0\zeta\delta_{ij}$, ${\partial_i\partial_j\zeta}/{\cH}$, and $\partial_k V^k\delta_{ij}$. 

When computing the effect of the coordinate change on the short-scale power spectrum it is straightforward to see that, 
since $\braket{\zeta(\eta,\vec{q}_1)\partial_0\zeta(\eta,\vec{q}_2)} = 
\braket{\partial_0\zeta(\eta,\vec{q}_2)\zeta(\eta,\vec{q}_1)}$ for $\vec{q}_1\neq\vec{q}_2$, 
only $\xi^0$ and $B^0_{ij}$ will contribute if we take the origin of spatial coordinates to lie at the middle point $(\vec{q}_1+\vec{q}_2)/2$. 
The final result is given by
\begin{equation}
\label{eq:app_D_intro-E}
\braket{\zeta^F\zeta^F}(\eta,r) = \braket{\zeta\zeta}(\eta,r) + \xi^0(\eta,\vec{0})\partial_0\braket{\zeta\zeta}(\eta,r) 
+ \frac{1}{4}B^0_{ij}(\eta)r^ir^j\partial_0\braket{\zeta\zeta}(\eta,r)\,\,,
\end{equation}
where we defined $\vec{r}\equiv\vec{q}_1-\vec{q}_2$, with $\abs{\vec{r}} = r$. 
Once we go to Fourier space and 
correlate \eq{app_D_intro-E} with the long mode, 
powers of $r^i$ turn into derivatives of the short-scale power spectrum with respect to $k_s$: 
therefore they will not play an important role in this discussion.\footnote{More precisely, we have that (schematically) 
\begin{equation*}
r^ir^j\to\frac{\partial^2}{\partial k_s^i\partial k_s^j}\sim\frac{k_s^ik_s^j}{k_s^{4}}\frac{\dif}{\dif\log k_s}\,\,,
\end{equation*}
so that in principle the term involving $B^0_{ij}(\eta)$ could correct the CFC bispectrum at order $k^2_\ell/k^2_s$. 
Therefore, it is important to check what is the time dependence of this term as well.} 
\eq{app_D_intro-F} tells us that, 
if we are interested in the time dependence of the final result, 
we need to compute how the two terms 
\begin{equation}
\label{eq:app_D_intro-F}
P_{\zeta\xi^0}(\eta,k_\ell)\partial_0P_{\zeta}(\eta,k_s)\,\,,\quad P_{\zeta B^0_{ij}}(\eta,k_\ell)\partial_0P_{\zeta}(\eta,k_s)\,\,,
\end{equation}
behave for $\eta\to 0^-$, with $P_{XY}(\eta,k)$ denoting the correlation function $\braket{X(\eta,\vec{k})Y(\eta,\vec{k}'}'$. 
The calculation is carried out in detail in Appendix \ref{app:appendix_D}: 
there it is shown how the two terms of \eq{app_D_intro-F} have a time dependence $\sim(-k\eta)^{{3}/{2}-\nu}$ as $\eta\to 0^-$, 
where $\nu\equiv i\mu$. Therefore as long as $m > 0$ (\ie $\nu<3/2$) they go to zero at late times, and do not affect the CFC bispectrum. 
If we take $\nu = {3}/{2}$, instead, $\sigma$ becomes massless and does not decay on super-Hubble scales. 
The mixing between $\zeta$ and $\sigma$ continues indefinitely after Hubble exit 
and the power spectrum of $\zeta$ does not go to a constant, 
but diverges as $\log(-k\eta)$ for $\eta\to 0^-$. 

Let us now briefly discuss how the bias expansion is augmented in presence of non-Gaussianity due to a massive spin-$0$ mediator. 
Recall that the spin gives an effect non-degenerate with the mass only through the Legendre polynomials 
$\mathcal{L}_s(\vers{k}_1\cdot\vers{k}_3)$, $\mathcal{L}_s(\vers{k}_2\cdot\vers{k}_3)$, 
and that halo bias is only sensitive to the monopole of the primordial bispectrum in the squeezed limit. Thus, we can again focus on zero spin.
At leading order in the squeezed limit, we see that the bispectrum of $\zeta$ takes the form\footnote{At this order in $k_\ell/k_s$, 
higher spins will show a scale dependence $\sim (k_\ell/k_s)^{3/2+s}$ \cite{Lee:2016vti,MoradinezhadDizgah:2017szk}. }
\begin{equation}
\label{eq:collider-A}
B_\zeta(k_1,k_2,k_3)\propto\cos\bigg(\mu\log\frac{k_\ell}{k_s}\bigg)\bigg(\frac{k_\ell}{k_s}\bigg)^{\frac{3}{2}}P_\zeta(k_s)P_\zeta(k_\ell)\,\,,
\end{equation}
where we have dropped the phase $\phi_0$ for simplicity since it does not play a role in this discussion. 
We can then see how the situation is very different from the one described at the beginning of Section \ref{sec:bias_resonant_NG}. 
In this case, the bispectrum shows \emph{logarithmic} oscillations with $k_\ell$, of frequency $\mu$: 
it cannot be represented by a series like that of \eq{bias_exp-D}, then, unless the parameter $\Delta$ is a complex number.
Indeed, we can rewrite \eq{collider-A} as
\begin{equation}
\label{eq:collider-B}
B_\zeta(k_1,k_2,k_3)\propto\bigg[\bigg(\frac{k_\ell}{k_s}\bigg)^{\frac{3}{2}+i\mu} 
+ \bigg(\frac{k_\ell}{k_s}\bigg)^{\frac{3}{2}-i\mu}\bigg]P_\zeta(k_s)P_\zeta(k_\ell)\,\,.
\end{equation}
Therefore we see that if we allow for complex scalings $\Delta$ 
it is possible to have full control on the bias expansion even if all the operators $\Psi^{(\Delta+2n)}$ are included, 
since all the oscillations are effectively resummed for any value of $\mu$. 
Correspondingly, it is still possible to derive the bias coefficients $b_{\Psi^{(\Delta+2n)}}$ using the response approach of \eqsII{bias_exp-E}{bias_exp-F}, 
and their specialization to a universal mass function given by \eq{bias_exp-K}, at any order in the squeezed limit. 

In the rest of this work, however, we will not use this approach: 
in order to better compare with the results of \cite{MoradinezhadDizgah:2017szk}, 
we use the resummation of \eq{bias_exp-P} to compute the non-Gaussian bias $\Delta b_1(k)$. 
By doing this, we are also resumming the contributions due to the transfer functions acting on the short-scale perturbations: 
indeed, \eq{bias_exp-J} assumes $\vec{k}_1\sim\vec{k}_2\sim\vec{k}_s$. 
As in Section \ref{sec:bias_resonant_NG}, we assume a Sheth-Tormen mass function and fix $z=0$. 
Given that, as we discussed above, the only effect of the transformation to CFC is that of removing the consistency relation, 
our analysis will be very similar to that of \cite{MoradinezhadDizgah:2017szk}. 
The main differences in our computation of $\Delta b_1$ are that we use the full bispectrum of \eq{QSF_bispectrum} 
(and not only its ultra-squeezed version with $k_1 = k_2 = k_s$), 
and we do not impose a cut-off $k_s/k_\ell\geq 10$ in the integral of \eq{bias_exp-Q}.

\subsection{Results}
\label{sec:QSF_results}

\noindent We consider the cases $\mu = 1$ and $\mu = 5$: 
for each value of $\mu$ we show both the case with $c_\pi = 1$, 
and the case with $c_\pi = \num{0.024}$ (that we call $c_\pi\ll 1$ in the plots), 
\ie the smallest value currently allowed from \emph{Planck} \cite{Ade:2015lrj}. 
The requirement of having a valid perturbative treatment of primordial non-Gaussianity implies $C_s < 1$ \cite{Lee:2016vti}. 
In our analysis, following \cite{MoradinezhadDizgah:2017szk}, we will take $C_s = 1$: 
the results can be quickly translated to different values of $C_s$ by a simple rescaling. 
Since for $\mu\gg 1$ the overall factor $f^{(0)}$ is 
\begin{equation}
\label{eq:asymptotic_overall_factor}
f^{(0)} = 
\begin{cases}
\text{${-\dfrac{\pi^{9/2}\mu^{3/2}e^{-{\pi\mu}}}{2}}$ for $c_\pi=1\,\,,$} \\[1em]
\text{${-\dfrac{\pi^4 e^{-\frac{\pi\mu}{2}}}{\sqrt{2}}}$ for $c_\pi\ll 1\,\,,$}
\end{cases}
\end{equation}
the signal for the $\mu = 5$ case will be much smaller than the $\mu = 1$ one. 
However, it is also important to stress that high values of $\mu$ lead to faster oscillations of the bispectrum with $\log(k_\ell/k_s)$, 
so that it is worth studying how they are imprinted on both the scale dependence of the halo bias and its dependence on the halo mass.

\begin{figure}[H]
\myfloatalign
\centering
\begin{tabular}{c}
\includegraphics[width=0.65\columnwidth]{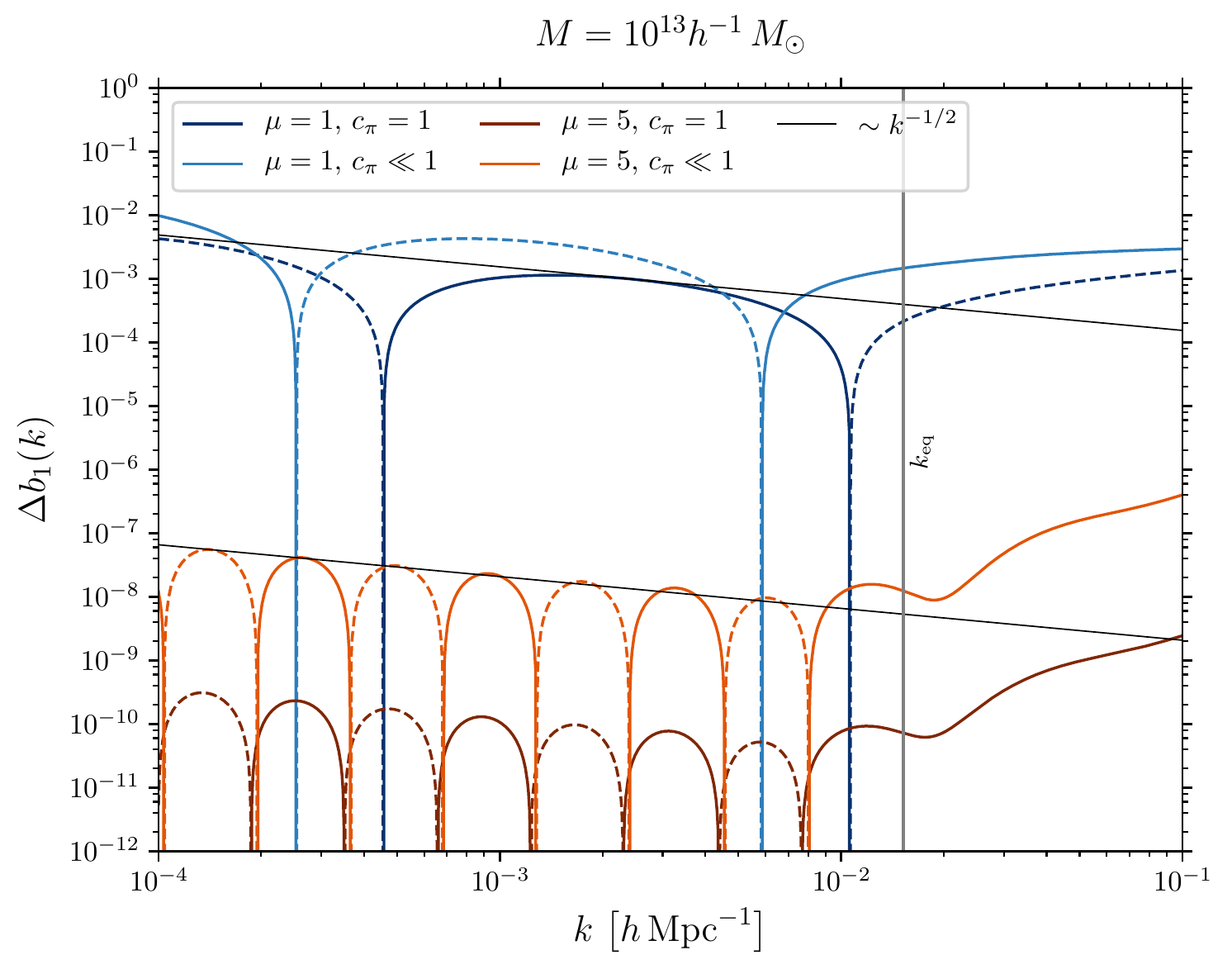} \\
\includegraphics[width=0.65\columnwidth]{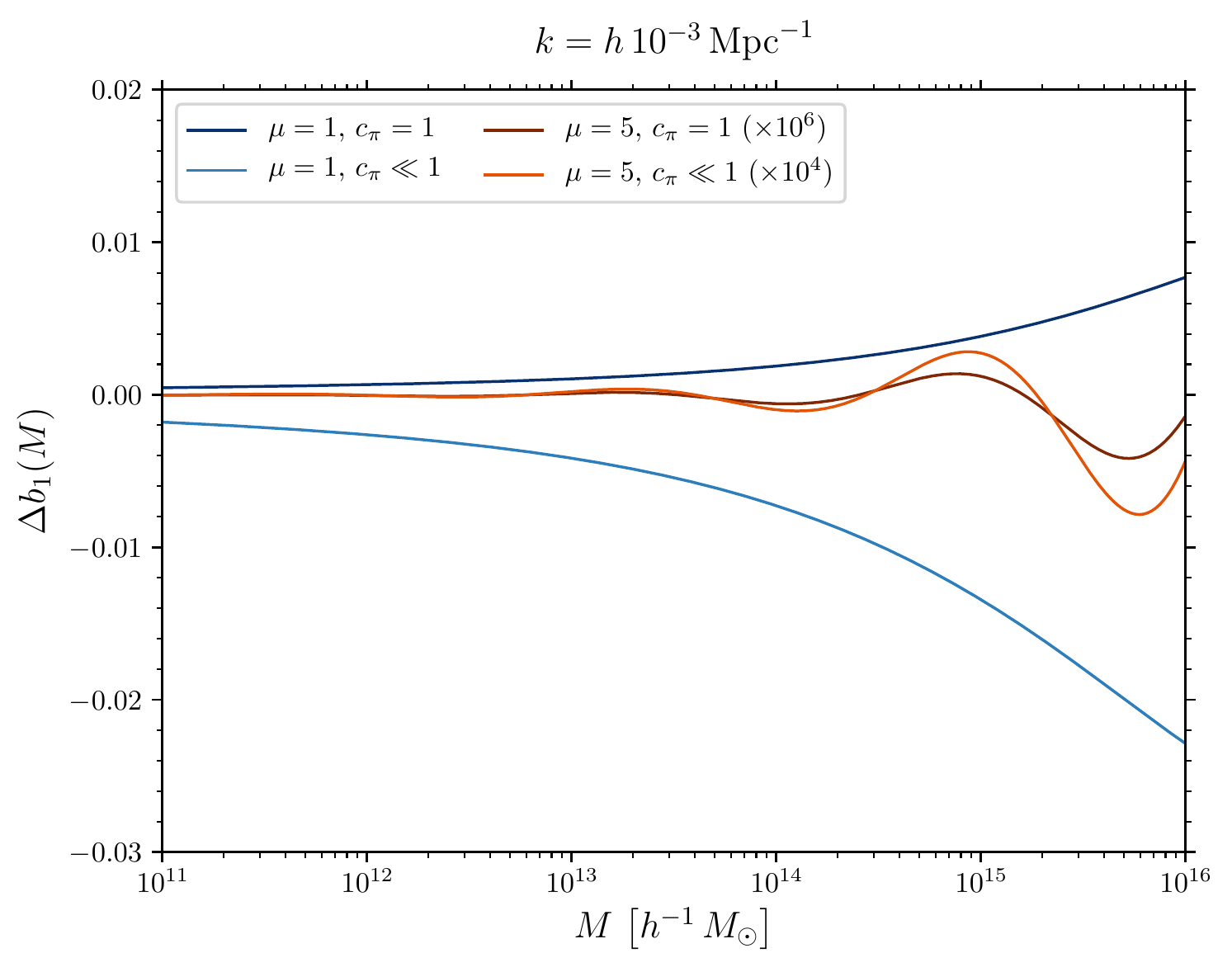}
\end{tabular}
\caption{Dependence of $\Delta b_1$ on $k$ at $M=\num{e13}{h^{-1}\,M_\odot}$ (top panel) and $M$ at $k = h\,\mpc{e-3}$ (bottom panel). 
We see that, as expected, $\Delta b_1\sim1/\sqrt{k}$ on large scales: 
due to the transfer function, departures from this scaling are seen at $k\sim k_\text{eq}$. 
While for $\mu = 1$ the dependence on the halo mass is very mild, 
for $\mu = 5$ oscillations with $M$ can be seen at large halo masses $M\gtrsim\num{e14}{h^{-1}\,M_\odot}$. }
\label{fig:QSF_kL_and_M}
\end{figure}

The top panel of Fig.~\ref{fig:QSF_kL_and_M} shows that the overall scale dependence of the non-Gaussian bias is $\sim 1/\sqrt{k}$, 
as we expect from \eq{QSF_bispectrum}. 
The oscillations with $k$, instead, take the form $\Delta b_1\sim\cos(\mu\log k)$. 
Oscillations with $M$ are also present: this is shown by the bottom panel of Fig.~\ref{fig:QSF_kL_and_M}. 
We see that such oscillations become more relevant for increasing halo mass and for increasing mass of the particles coupled to the inflaton, 
thereby offering an alternative way to constrain these heavy fields. 

Finally, we comment on the impact of imposing a cut-off $k_s/k\geq 10$ in the integral of \eq{bias_exp-Q}. 
In this case, we take the same values for $M$ and $\mu$ as \cite{MoradinezhadDizgah:2017szk}, 
\ie $M=\num{3e13}{h^{-1}\,M_\odot}$, $\mu=\sqrt{3}/2$. 
We find that the relative difference between the calculation of $\Delta b_1(k)$ with and without cutoff is an increasing function of $k$. 
While it is smaller than \num{d-3} for $k\approx\mpc{e-4}$, it becomes of order $1$ for $k\approx\mpc{e-2}$.

\section{Conclusions and discussion}
\label{sec:conclusions}

\noindent In this paper, we discussed the scale-dependent contribution to halo clustering from resonant non-Gaussianity, 
which arises from string theory-motivated models like axion monodromy. 
Working in Conformal Fermi Coordinates, we emphasized the importance of accounting for gauge artifacts: 
we find that resonant non-Gaussianity makes predictions for the scale dependence of the bias that are very different from the ones computed in \cite{CyrRacine:2011rx}, 
where the unphysical contribution from the consistency relation was taken into account. 
We see that $\Delta b_1$ oscillates strongly with scale, with an envelope similar to that of equilateral non-Gaussianity. 
More precisely, the envelope of the oscillations is equivalent to having $\fnleq\approx\num{0.6}\fnlres/\alpha^{3/2}$. 

Since increasing $\alpha$ leads to faster oscillations with both $k$ and halo mass, 
the scale-dependent bias could in principle offer a window on a region of parameter space where CMB constraints 
(that come mainly from the power spectrum) are weaker.
However, we show that measurements of the galaxy power spectrum from upcoming surveys 
will not be able to detect the non-Gaussian bias, even for optimistic values $\alpha = \mathcal{O}(\num{d3})$. 

As a second example of primordial bispectrum with oscillatory features, 
we considered the case of ``cosmological collider'' models, 
in which the inflaton is coupled with massive particles of spin $s$ and mass $m$. 
We have shown that, also in these models, the only effect of switching to Conformal Fermi Coordinates after averaging over angles is that 
of removing the $\mathcal{O}(k_\ell^0/k_s^0)$ term in the squeezed limit (\ie the consistency relation). 
Therefore, our conclusions for the scale dependence of the bias agree with the previous literature 
like, for example, \cite{MoradinezhadDizgah:2017szk} 
(that carried out detailed forecasts for the $s=0$, $s=1$ and $s=2$ case). 
Finally we show that, analogously to what happens in resonant non-Gaussianity, also in these models the bias oscillates with the halo mass, 
with oscillations becoming stronger with increasing particle mass $m$. 

As a byproduct of this work, 
we have derived the link between the short-scale power spectrum of $\zeta$ in the background of the long mode during inflation, 
and that of the Newtonian potentials $\phi$ and $\psi$ at early times during the Hot Big Bang phase, in the background of the same long mode. 
This makes use of Conformal Fermi Coordinates \cite{Dai:2015rda,Dai:2015jaa}, 
and has been obtained for an isotropic long mode: 
in this case, CFC make explicit that the short modes evolve in a separate FLRW universe, 
and considering an isotropic long mode is enough to compute all the LIMD 
(``local-in-matter-density'') biases \cite{Desjacques:2016bnm}. 
However, it would be interesting to provide the connection between $\zeta$ and $\{\psi,\phi\}$ 
also in the case where large-scale tidal fields are present at least for two reasons. 
First, it would allow to generalize the initial conditions for separate universe simulations \cite{Baldauf:2011bh,Wagner:2014aka,Baldauf:2015vio}, 
which have recently begun to include large-scale tidal fields and not only modifications to the homogeneous and isotropic FLRW \cite{Schmidt:2018hbj}, 
to include an anisotropy in the initial power spectrum coming from mode coupling during inflation. 
Then, as discussed in the main text, we know that if massive higher-spin fields are coupled to the inflaton 
the bispectrum of $\zeta$ acquires a peculiar dependence on $\vers{k}_s\cdot\vers{k}_\ell$ in the squeezed limit, 
depending on the spin of the massive particle \cite{Arkani-Hamed:2015bza,Lee:2016vti}. 
By considering an isotropic long mode we are effectively averaging over $\vers{k}_s\cdot\vers{k}_\ell$, 
so that we are oblivious to such effects, which can for example affect galaxy shapes (as shown in \cite{Schmidt:2015xka,Chisari:2016xki}) and, 
in general, the galaxy bispectrum \cite{Assassi:2015fma,MoradinezhadDizgah:2018ssw}. 
While these effects coming from the ``non-linear'' relation between $\zeta$ and $\{\psi,\phi\}$ will be analytic in $k_\ell/k_s$, 
characterizing them can still be important if one wants to carry out forecasts for future surveys. 
We leave this to future work.

\section*{Acknowledgements}

\noindent It is a pleasure to thank Lorenzo Bordin, Paolo Creminelli, Dionysios Karagiannis, 
Titouan Lazeyras, Hayden Lee, Michele Liguori and Marko Simonovi\'{c} for useful discussions. 
G. C. and F. S. acknowledge support from the Starting Grant (ERC-2015- STG 678652) “GrInflaGal” from the European Research Council. 
E. P. is supported by the Delta-ITP consortium, a program of the Netherlands Organization for Scientific Research (NWO) that is funded by the Dutch 
Ministry of Education, Culture and Science (OCW). 
This work is part of the research programme VIDI with Project No. 680-47-535, 
which is (partly) financed by the Netherlands Organisation for Scientific Research (NWO).

\appendix

\section{Initial conditions from CFC: single-clock inflation}
\label{app:appendix_A}

\noindent In this appendix we describe how to match the calculation of \cite{Cabass:2016cgp} to that of \cite{Dai:2015jaa}. 
We consider an isotropic long mode, so that the CFC background metric is that of a curved FLRW universe. 
Since we can follow the CFC patch from Hubble exit to Hubble re-entry of the short modes, 
we just need the \emph{linear} relation between $\zeta_s^F$ during inflation (of which we know the power spectrum) 
and, say, some quantity like the Newtonian potentials $\phi^F_s$, $\psi^F_s$. For the sake of simplicity, 
in the rest of this appendix we will drop the superscript $F$ and the subscript $s$ and we work in Planck units $8\pi G_{\rm N} = \mpl^{-2} = 1$. 

Let us be more precise: following \cite{Dai:2015jaa}, the Newtonian potentials in CFC are defined by 
\begin{equation}
\label{eq:app_A-red-1}
\dif s^2 = a^2\bigg[{-(1+2\phi)\dif\eta^2} + \frac{(1-2\psi)\dif\vec{x}^2}{(1 + K\abs{\vec{x}}^2/4)^2}\bigg]\,\,.
\end{equation}
The equations for the evolution of $\phi$, $\psi$ and the small-scale density and velocity perturbations ($\delta$ and $v^i$) 
for non-relativistic pressure-free matter have been also derived in \cite{Dai:2015jaa} (see Eqs.~(5.8) there). 
In order to solve these equations the initial conditions for $\phi$, $\psi$, $\delta$ and $v^i$ at early times $\eta\to 0$ are needed. 
The initial density and velocity perturbations can be expressed in terms of the initial Newtonian potentials, 
so it is enough to know $\phi|_\text{ini}$ and $\psi|_\text{ini}$: 
these contain the coupling to $K$ that comes from the inflationary dynamics (encoded in the CFC power spectrum of $\zeta$). 
In the following, we compute these initial conditions. 

Recall that, in a curved FLRW background, 
the curvature perturbation\footnote{In this appendix we are using a potentially confusing notation: 
we indicate by $\zeta$ the curvature perturbations on \textit{comoving} hyperslices, as in \cite{Maldacena:2002vr}, 
while we use $\zeta_{\text{ud}}$ for curvature perturbations on \textit{uniform density} slices. } 
on constant energy hypersurfaces, \emph{i.e.}
\begin{equation}
\label{eq:app_A-A}
\zeta_{\text{ud}} \equiv A - \frac{\cH\delta\rho}{\rho_0'}\,\,,
\end{equation}
is gauge-invariant at linear order in perturbations and is conserved outside the sound horizon. 
Here $A$, in a general gauge, is defined by \cite{Wands:2000dp}
\begin{equation}
\label{eq:app_A-B}
g_{ij} = \frac{a^2(1 + 2A)\delta_{ij}}{\Big(1 + \frac{K\abs{\vec{x}}^2}{4}\Big)^2} + 2\hat{\partial}_i\hat{\partial}_j B\,\,,
\end{equation}
where we have considered only scalar perturbations and $\hat{\partial}_i$ denotes the covariant derivative on a three-sphere/three-hyperboloid. 
$\rho_0$ and $\delta\rho$, instead, are defined from the stress-energy tensor as $T_0^{\hphantom{0}0} = -(\rho_0 + \delta\rho)$. 

The calculation of the CFC power spectrum for the short modes during inflation is done in unitary gauge, \ie $B=0$ and an unperturbed inflaton ($\delta\phi=0$). 
Let us consider for simplicity the case of minimal slow-roll inflation, 
where higher-derivative operators in the inflaton+gravity action are turned off. 
The same conclusions will apply in general to models that are described by the EFT of Inflation \cite{Cheung:2007st}. 
By its fully non-linear definition, $\zeta$ is simply $A$ in unitary gauge. 
Moreover, $\zeta$ is conserved when all the modes exit the Hubble radius during inflation \cite{Maldacena:2002vr}, 
and its power spectrum (and the coupling with the CFC curvature $K$) at late times is known \cite{Cabass:2016cgp}. 
$T_0^{\hphantom{0}0}$, instead, is equal to 
\begin{equation}
\label{eq:app_A-C}
T_0^{\hphantom{0}0}=-(\rho_0 + \delta\rho)={-\bigg(\frac{3K}{a^2}+\frac{3\cH^2}{a^2} - \frac{2\eps\cH^2\delta N}{a^2}\bigg)}\,\,,
\end{equation}
where $g_{00} = {-a^2}(1+2\delta N)$, $\eps=-\dot{H}/H^2$ and we work at leading order in curvature. 
The solution for $\delta N$ comes from the shift constraint equation: 
its expression at leading order in $K$ is given by
\begin{equation}
\label{eq:app_A-D}
\delta N = \frac{\zeta'}{\cH} + \frac{K\chi}{\cH}\,\,,
\end{equation}
where $\partial_i\chi$ is defined in \eq{maldacena_gauge-C-2}. 
It is then straightforward to see that \eq{app_A-A} becomes
\begin{equation}
\label{eq:app_A-E}
\zeta_{\text{ud}} = \zeta - \frac{\eps\cH^2\delta N}{3(K + \eps\cH^2)}\,\,,
\end{equation}
which goes to $\zeta$ for super-Hubble modes (\emph{i.e.} for $\eta\to 0^{+}$ in a decelerated universe). 

We can now compute $\phi|_\text{ini}$, $\psi|_\text{ini}$ by working in Newtonian gauge, \ie using \eq{app_A-red-1}. 
The lapse constraint equation for a perfect fluid, 
at first order in perturbations on a curved FLRW background, reads
\begin{equation}
\label{eq:app_A-G}
\frac{2\hat{\partial}^2\psi}{a^2} + \frac{6K\psi}{a^2} - \frac{6\cH\psi'}{a^2} - \frac{6\cH^2\phi}{a^2} = \delta\rho\,\,,
\end{equation}
where we have again stopped at first order in the curvature $K$ and by $\hat{\partial}^2$ we indicate the Laplacian on the three-sphere/three-hyperboloid. 
This allows to express $\delta\rho$ in terms of $\psi = \phi$, so that \eq{app_A-A} becomes
\begin{equation}
\label{eq:app_A-H}
\zeta_{\text{ud}} = -\psi + \frac{\cH\hat{\partial}^2\psi + 3\cH K\psi - 3\cH^2\psi' - 3\cH^3\psi}{3K\cH + 3\eps\cH^3}\,\,.
\end{equation}
On super-Hubble scales ($\eta\to 0^{+}$ for a decelerated universe), 
${- 3\cH^3\psi}$ and $3\eps\cH^3$ dominate in the second term on the right-hand side of the above equation, and we arrive at 
\begin{equation}
\label{eq:app_A-I}
(\zeta_{\text{ud}})|_\text{super-Hubble} = {-\bigg(1+\frac{1}{\eps}\bigg)}\psi = -\frac{5+3w}{3(1+w)}\psi\,\,,
\end{equation}
\ie we find \eq{initial_conditions}. At early times, the corrections of the long mode to the expansion history go to zero \cite{Dai:2015jaa,Cabass:2016cgp},\footnote{See 
Eq.~(3.18) of \cite{Dai:2015jaa} and Appendix A of \cite{Cabass:2016cgp} for a more detailed discussion.} 
so no additional coupling between the long and the short modes will come from $w$. 
That is, we can simply take $w=1/3$ if the short modes re-enter the Hubble radius during radiation dominance, 
or $w=0$ if they re-enter during matter dominance. More precisely, taking $w=0$ gives the initial conditions for Eqs.~(5.8) of \cite{Dai:2015jaa}.

\section{Details on the resummation of non-Gaussian biases}
\label{app:resummation_proof}

\noindent In this appendix we present a more detailed proof of \eq{bias_formula-C}. 
First, let us expand \eq{local_PS-B} to higher orders in the squeezed limit. 
Using \eqsII{bias_exp-A}{bias_exp-D}, we find 
\begin{equation}
\label{eq:resummation_eq-A}
\begin{split}
{\cal F}^{(3)}_\ast(k) &= \frac{1}{4\sigma_\ast^2P_\phi(k)}\int\frac{\dif^3 k_s}{(2\pi)^3}\mathcal{M}_\ast(k_1)\mathcal{M}_\ast(k_2)B_\phi(k_1,k_2,k) \\
&\approx\frac{1}{4\sigma_\ast^2P_\phi(k)}\int\frac{\dif^3 k_s}{(2\pi)^3}\mathcal{M}_\ast^2(k_s)\bigg[
4\sum_{n=0}^{+\infty}a_{0,2n}\bigg(\frac{k}{k_s}\bigg)^{\Delta+2n} P_\phi(k_s)P_\phi(k)\bigg] \\
&=\sum_{n=0}^{+\infty}\frac{a_{0,2n}k^{\Delta+2n}}{\sigma_\ast^2}\int\frac{\dif^3 k_s}{(2\pi)^3}\frac{W^2_\ast(k_s)P_\text{m}(k_s)}{k_s^{\Delta+2n}} \\
&=\sum_{n=0}^{+\infty}a_{0,2n}k^{\Delta+2n}\underbrace{\frac{\sigma^2_{\ast,-\Delta/2-n}}{\sigma_\ast^2}}_{\hphantom{y_n\,}\equiv\,y_n}\,\,,
\end{split}
\end{equation}
where we defined $\sigma^2_{\ast,-\Delta/2-n}/\sigma_\ast^2\equiv y_n$ for simplicity (since it will appear often in the following computations). 
Then, let us also define 
\begin{equation}
\label{eq:resummation_eq-B}
x_n\equiv\frac{\partial\log\sigma^2_{\ast,-\Delta/2-n}}{\partial\log\sigma^2_\ast} = \frac{\partial\log(\sigma^2_\ast y_n)}{\partial\log\sigma^2_\ast}\,\,.
\end{equation}
With this definition, we have that \eq{bias_exp-K} (and its generalization to arbitrary $n$) becomes
\begin{equation}
\label{eq:resummation_eq-C}
b_{\Psi^{(\Delta+2n)}} = \big[b_{\Psi^{(0)}} + 4(x_n-1)\big]y_n\,\,,
\end{equation}
where $b_{\Psi^{(0)}}=b_\phi$. Consequently, the left-hand side of \eq{bias_formula-C} becomes
\begin{equation}
\label{eq:resummation_eq-D}
\Delta b_1(k)\mathcal{M}(k) = \sum_{n=0}^{+\infty}a_{0,2n}b_{\Psi^{(\Delta+2n)}}k^{\Delta+2n} = 
\sum_{n=0}^{+\infty}a_{0,2n}\big[b_{\Psi^{(0)}} + 4(x_n-1)\big]y_nk^{\Delta+2n}\,\,.
\end{equation}

Let us start by neglecting the terms coming from the Jacobian, \ie putting $x_n=0$. 
We have to compare \eq{resummation_eq-D} with the right-hand side of \eq{bias_formula-C}, \ie
\begin{equation}
\label{eq:resummation_eq-E}
\Delta b_1(k)\mathcal{M}(k) = \bigg[b_{\Psi^{(0)}} + 4\bigg(\frac{\partial\log\big(\sigma^2_\ast{\cal F}^{(3)}_\ast(k)\big)}{\partial\log\sigma_\ast^2}-1\bigg)\bigg]{\cal F}^{(3)}_\ast(k)\,\,.
\end{equation}
To do this, using \eq{resummation_eq-A} we write
\begin{equation}
\label{eq:resummation_eq-F}
\begin{split}
\frac{\partial\log\big(\sigma^2_\ast{\cal F}^{(3)}_\ast(k)\big)}{\partial\log\sigma_\ast^2} - 1 &= 
\frac{1}{{\cal F}^{(3)}_\ast(k)}\frac{\partial{\cal F}^{(3)}_\ast(k)}{\partial\log\sigma^2_\ast} \\
&=\frac{1}{{\cal F}^{(3)}_\ast(k)}\bigg[\sum_{n=0}^{+\infty}a_{0,2n}k^{\Delta+2n}\frac{\partial y_n}{\partial\log\sigma^2_\ast}\bigg] \\
&=\frac{1}{{\cal F}^{(3)}_\ast(k)}\bigg[\sum_{n=0}^{+\infty}a_{0,2n}k^{\Delta+2n}y_n\bigg(\frac{\partial\log(\sigma^2_\ast y_n)}{\partial\log\sigma^2_\ast}-1\bigg)\bigg] \\
&=\frac{1}{{\cal F}^{(3)}_\ast(k)}\bigg[\sum_{n=0}^{+\infty}a_{0,2n}k^{\Delta+2n}y_n(x_n-1)\bigg]\,\,.
\end{split}\end{equation}
If we assume $x_n=0$, using \eq{resummation_eq-A} we see that this is equal to
\begin{equation}
\label{eq:resummation_eq-G}
\frac{\partial\log\big(\sigma^2_\ast{\cal F}^{(3)}_\ast(k)\big)}{\partial\log\sigma_\ast^2} - 1 = -\frac{1}{{\cal F}^{(3)}_\ast(k)}\sum_{n=0}^{+\infty}a_{0,2n}k^{\Delta+2n}y_n = -1\,\,,
\end{equation}
so that \eq{resummation_eq-E} becomes 
\begin{equation}
\label{eq:resummation_eq-H}
\Delta b_1(k)\mathcal{M}(k) = (b_{\Psi^{(0)}} - 4){\cal F}^{(3)}_\ast(k) = 
\sum_{n=0}^{+\infty}a_{0,2n}(b_{\Psi^{(0)}} - 4)y_n k^{\Delta+2n}\,\,.
\end{equation}
Using \eqsII{resummation_eq-C}{resummation_eq-D} with $x_n = 0$, we see that \eq{bias_formula-C} holds. 

Things are more complicated if we do not assume $x_n=0$. \eq{bias_formula-C} must be checked 
order by order in $k$. If we stop at $n=0$, the agreement is trivial. Let us go up to $n=1$. Then, \eq{resummation_eq-F} becomes
\begin{equation}
\label{eq:resummation_eq-I}
\begin{split}
\frac{\partial\log\big(\sigma^2_\ast{\cal F}^{(3)}_\ast(k)\big)}{\partial\log\sigma_\ast^2} - 1 &= \frac{1}{{\cal F}^{(3)}_\ast(k)}\bigg[\sum_{n=0}^{1}a_{0,2n}k^{\Delta+2n}y_n(x_n-1) \\
&= \frac{a_{0,0}k^{\Delta}y_0(x_0-1) + a_{0,2}k^{\Delta+2}y_2(x_2-1)}{a_{0,0}k^{\Delta}y_0 + a_{0,2}k^{\Delta+2}y_2} \\
&= (x_0-1) - (x_0-1)\frac{a_{0,2}}{a_{0,0}}\frac{y_2}{y_0}k^2 + (x_2-1)\frac{a_{0,2}}{a_{0,0}}\frac{y_2}{y_0}k^2 + \mathcal{O}(k^4)\,\,,
\end{split}
\end{equation}
where we have expanded up to the relevant order, \ie $k^2$. 
Therefore, the right-hand side of \eq{resummation_eq-E} becomes
\begin{equation}
\label{eq:resummation_eq-J}
\begin{split}
\text{r.h.s. of \eq{resummation_eq-E}} &= 
b_{\Psi^{(0)}}(a_{0,0}k^{\Delta}y_0 + a_{0,2}k^{\Delta+2}y_2) \\
&\;\;\;\; + 4\bigg[(x_0-1) - (x_0-1)\frac{a_{0,2}}{a_{0,0}}\frac{y_2}{y_0}k^2 + (x_2-1)\frac{a_{0,2}}{a_{0,0}}\frac{y_2}{y_0}k^2\bigg] \\
&\;\;\;\;\times (a_{0,0}k^{\Delta}y_0 + a_{0,2}k^{\Delta+2}y_2)
+\mathcal{O}(k^{\Delta+4}) \\
&= b_{\Psi^{(0)}}(a_{0,0}k^{\Delta}y_0 + a_{0,2}k^{\Delta+2}y_2) \\
&\;\;\;\;+ 4\big[(x_0-1)a_{0,0}k^\Delta y_0+(x_2-1)a_{0,2}k^{\Delta+2}y_2 + \mathcal{O}(k^{\Delta+4})\big] \\
&= a_{0,0}\big[b_{\Psi^{(0)}}+4(x_0-1)\big]y_0k^\Delta \\
&\;\;\;\;+ a_{0,2}\big[b_{\Psi^{(0)}}+4(x_2-1)\big]y_2k^{\Delta+2}+\mathcal{O}(k^{\Delta+4}) \\
&= \underbrace{a_{0,0} b_{\Psi^{(\Delta)}} k^\Delta + a_{0,2} b_{\Psi^{(\Delta+2)}} k^{\Delta+2}+\mathcal{O}(k^{\Delta+4})}_{
\text{equal to \eq{resummation_eq-D} at this order}}\,\,.
\end{split}
\end{equation}

It is then clear from the calculations shown here that \eq{bias_formula-C} can be proven at all orders in $k$. 
A proof can either proceed by induction, or by expanding $1/\mathcal{F}^{(3)}_\ast(k)$ in \eq{resummation_eq-I} as a power series,\footnote{More precisely, 
given a power series $f=\sum_{n=0}^{+\infty}a_nx^n$, 
one can recursively find an expression for the coefficients $b_n$ of the series $1/f = \sum_{n=0}^{+\infty}b_n x^n$. } 
and then doing a Cauchy product of this series with that for $\partial{\cal F}^{(3)}_\ast(k)\big)/\partial\log\sigma_\ast^2$, 
thereby finding an expression for $\partial\log\big(\sigma^2_\ast{\cal F}^{(3)}_\ast(k)\big)/\partial\log\sigma_\ast^2 - 1$ at all orders.

\section{Dependence of bias on frequency \texorpdfstring{$\alpha$}{\textbackslash alpha}}
\label{app:appendix_B}

\begin{figure}[H]
\myfloatalign
\centering
\begin{tabular}{c}
\includegraphics[width=0.65\columnwidth]{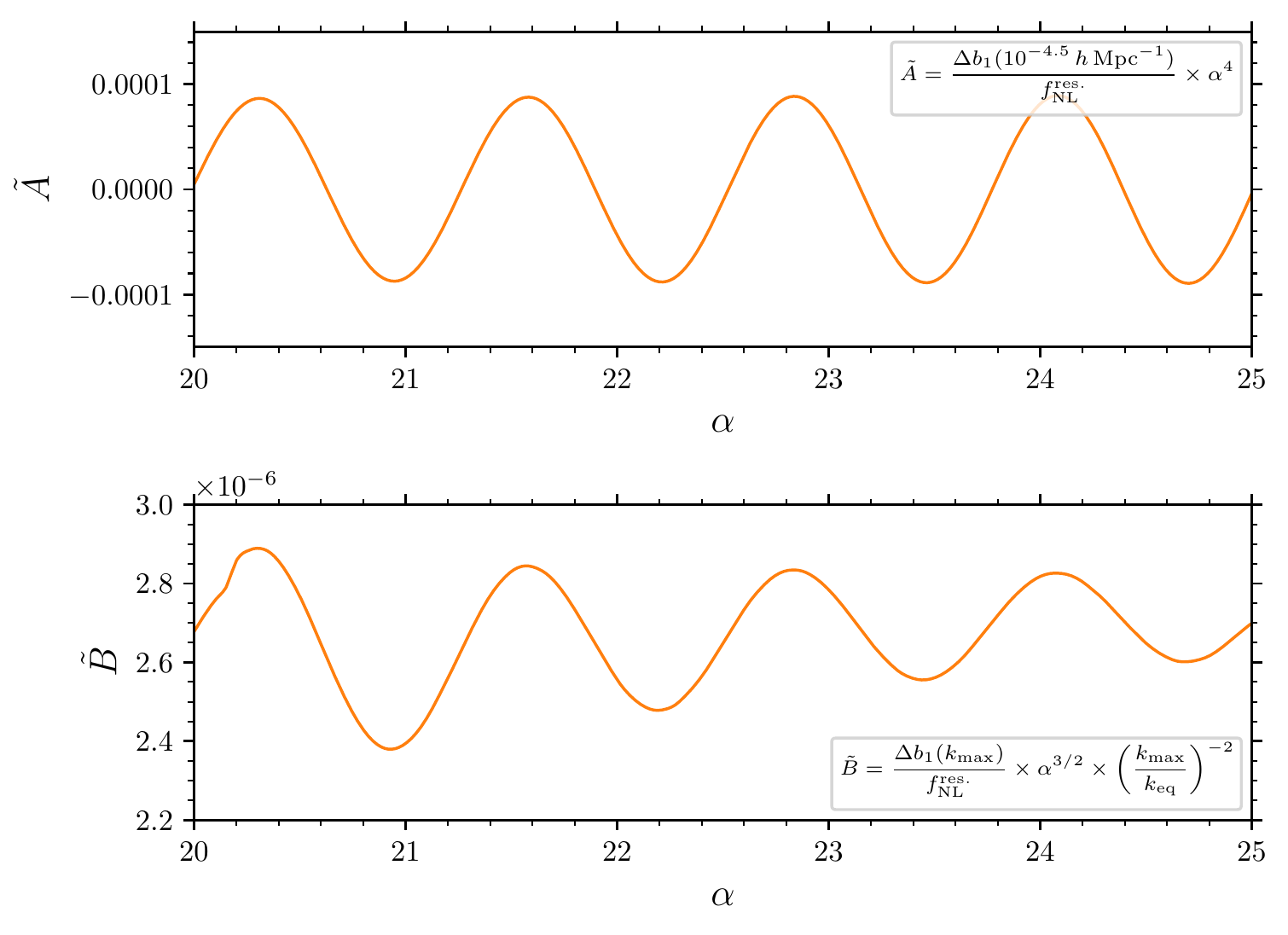} \\
\includegraphics[width=0.65\columnwidth]{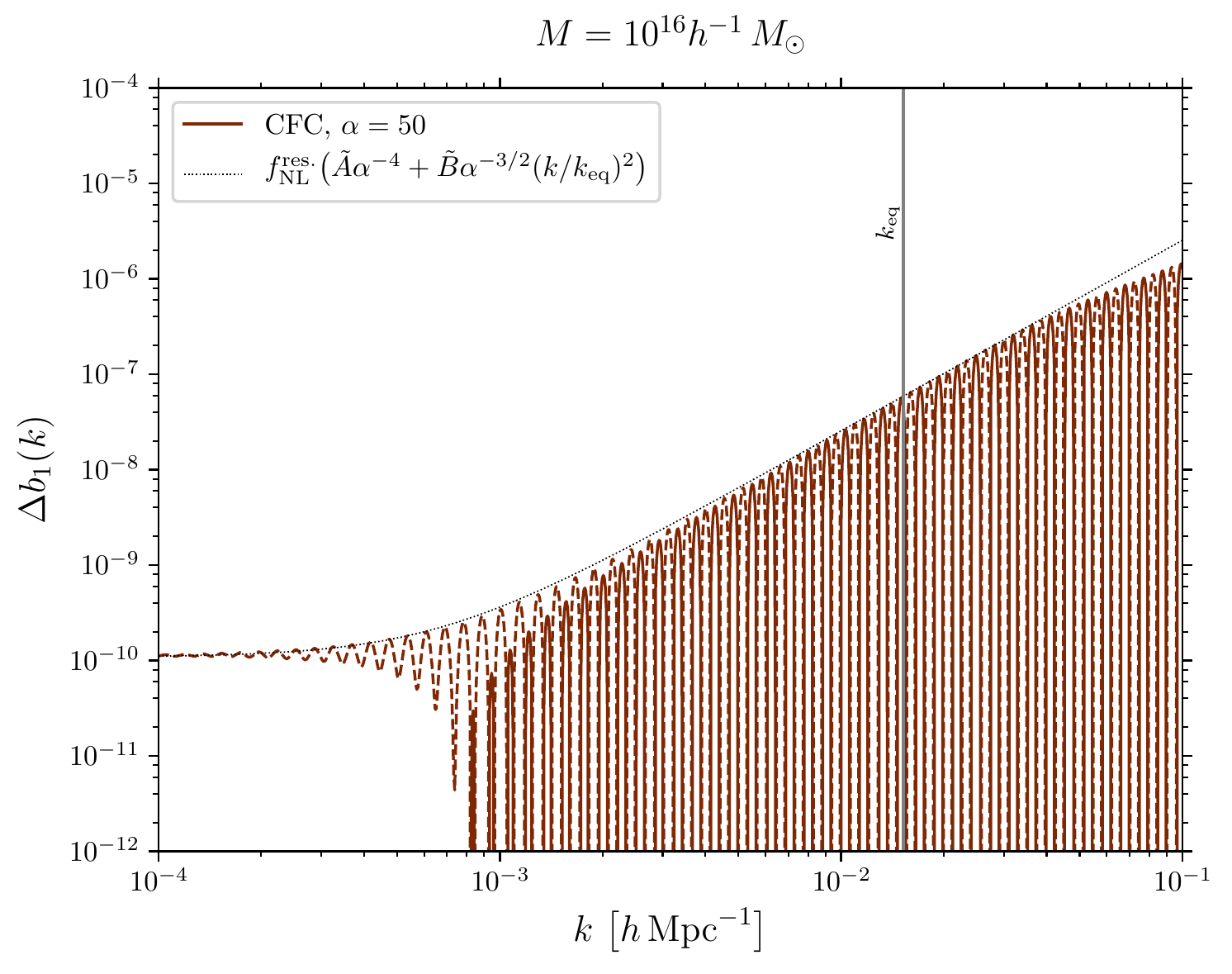}
\end{tabular}
\caption{Top panel: dependence of the parameters $\tilde{A} = A\,\alpha^4$ and $\tilde{B} = B\,\alpha^{3/2}$ of \eq{app_B_eq-A} 
on the dimensionless frequency of oscillations $\alpha$. 
Bottom panel: dependence of the non-Gaussian bias on $k$ for $\alpha=50$ and $M = \num{e16}{h^{-1}\,M_\odot}$ 
(the dashed lines represent negative values of $\Delta b_1$). 
We see that the conclusions drawn in Section \ref{sec:resonant_results} are confirmed. 
On short scales the bias oscillates logarithmically with $k$ and with a frequency $\propto\alpha$. 
For $k/k_\text{eq}\approx 10$ the transfer function leads to a deviation from the simple scaling $\Delta b_1\propto k^2$ of \eq{app_B_eq-A}. }
\label{fig:additional_plots}
\end{figure}

\noindent In this appendix we describe how we obtained the fit of \eq{fit} for the envelope of the oscillations of $\Delta b_1$ with $k$. 
We assume that the envelope consists of a constant and a term quadratic in $k$, \ie
\begin{equation}
\label{eq:app_B_eq-A}
{\Delta b_1(k)}=\fnlres\bigg[A+B\bigg(\frac{k}{k_\text{eq}}\bigg)^2\bigg]\,\,.
\end{equation}
The parameter $A$ is fitted from the scale-independent value of $\Delta b_1$ on large scales, where the term $\propto k^2$ is negligible. 
That is, we take $k=h\,\mpc{e-4.5}$ and compute the non-Gaussian bias there, neglecting the contribution from $B({k}/{k_\text{eq}})^2$ in \eq{app_B_eq-A}. 
We see that $A=\tilde{A}/\alpha^4$, where $\tilde{A}$ oscillates with $\alpha$ around zero: this is shown in the first plot of the 
top panel of Fig.~\ref{fig:additional_plots}. The parameter $B$ is obtained in a similar way: 
we select the local maximum $k_\text{max}$ around $k=h\,\mpc{e-2}$, and compute $B$ as
\begin{equation}
\label{eq:app_B_eq-B}
B=\frac{\Delta b_1(k_\text{max})}{\fnlres}\bigg(\frac{k_\text{max}}{k_\text{eq}}\bigg)^{-2}\,\,.
\end{equation}
We find that $B=\tilde{B}/\alpha^{3/2}$, where $\tilde{B}$ oscillates with $\alpha$ around $\num{2.7d-6}$ 
(see the second plot of the top panel of Fig.~\ref{fig:additional_plots}).

\section{Initial conditions from CFC: cosmological collider case}
\label{app:appendix_D}

\noindent In this appendix we show how to generalize the results of Section \ref{sec:bias_and_curvature} 
and Appendix \ref{app:appendix_A} 
to the case where additional fields, possibly with non-zero spin, 
are present during inflation. 
As discussed in the main text, 
we restrict to the case where the masses of these heavy fields are non-zero: 
the curvature perturbation, then, freezes on super-Hubble scales 
and we can use \eq{app_A-I}, 
provided that we compute the squeezed CFC bispectrum. 
The only non-trivial contribution to the bispectrum comes from the change in the time coordinate, 
that gives rise to terms containing the time derivative of the short-scale power spectrum: 
in this appendix we show how these terms vanish at late times as in the single-field case. 

As a prototype for generic models 
where the inflaton is coupled to some additional massive degrees of freedom at the quadratic level 
(so that the power spectrum of $\zeta$ gets corrections depending on the quadratic mixing between the inflaton and the massive field), 
we consider the leading in derivatives quadratic mixing in the EFT of Inflation, 
\ie $\mathcal{L}\supset\delta g^{00}\sigma$, 
where $\sigma$ is a massive minimally coupled scalar field. 
Higher spin fields can in general contribute to the quadratic Lagrangian 
also at zeroth order in derivatives. 
For example, consider a spin-$1$ field $\sigma_{\mu}$: 
it can contribute to the mixing via $\mathcal{L}\supset \delta g^{00}\sigma^0$. 
However, the spin of the field does not have an important effect on the power spectrum of $\zeta$, 
so we will focus on the spin-$0$ case only. 
We work in the decoupling limit, 
and then neglect metric perturbations in the action for the Goldstone boson $\pi$ of broken time diffeomorphisms. 
The quadratic action for $\pi$ and $\sigma$ then becomes
\begin{equation}
\label{eq:app_D-A}
\begin{split}
&S_{(2)} = \mpl^2\int\dif^4x\,\sqrt{-g}\,\dot{H}(\partial_\mu\pi)^2 
- \frac{1}{2}\int\dif^4x\,\sqrt{-g}\,\big((\partial_\mu\sigma)^2 + m^2\sigma^2\big) 
+ \omega_0^3\int\dif^4x\,\sqrt{-g}\,\dot{\pi}\sigma\,\,.
\end{split}
\end{equation}
Defining the scale $f_\pi^4 = 2\mpl^2\abs{\dot{H}} = -2\mpl^2\dot{H}$ 
(at which time translations are spontaneously broken and a description in terms of a Goldstone boson becomes applicable), 
and $\rho\equiv\omega^3_0f^{-2}_\pi$, 
we see that the action is that of two minimally coupled scalar fields $\sigma$ (massive) and $\pic\equiv f^2_\pi\pi$ (massless) in de Sitter, 
with interaction Lagrangian $\int\dif^3x\,\sqrt{-g}\,\mathcal{L}_\text{int} = \rho\int\dif^3x\,a^3\pi'\sigma$. 

To see how fast $\zeta = -H\pi$ evolves with time, 
we use the tree-level in-in formula with $H_\text{int} = -\int\dif^3x\,\sqrt{-g}\,\mathcal{L}_\text{int} = -\rho\int\dif^3x\,a^3\pi'\sigma$ 
to compute the power spectrum of $\pi$ at a finite time $\eta\to 0^-$. 
The general expression for the expectation value $\braket{\text{in}|O(\eta)|\text{in}}\equiv\braket{O(\eta)}$ 
of a generic operator $O(\eta;\vec{k}_1,\vec{k}_2,\dots)$ is given by \cite{Weinberg:2005vy}
\begin{equation}
\label{eq:app_D-B}
\braket{O(\eta)} = \braket{0|\;\widebar{T}e^{i\int_{-\infty^{-}}^\eta\dif s\, H^\text{I}(s)}\;O^\text{I}(\eta)\;Te^{-i\int_{-\infty^{+}}^\eta\dif s\, H^\text{I}(s)}\;|0}\,\,,
\end{equation}
where $\ket{0}$ is the free vacuum, the superscript $\text{I}$ denotes interaction picture operators, and $-\infty^{\pm} = -\infty(1\pm i\epsilon)$. 
For the action of \eq{app_D-A}, 
at tree level there is only one diagram that contributes to the two-point function of $\pic$, 
and the power spectrum of $\zeta$ takes the form \cite{Chen:2009zp,Chen:2012ge} 
(see also \cite{Tong:2017iat} for a non-perturbative calculation) 
\begin{equation}
\label{eq:app_D-C}
P_\zeta(\eta) = P_\zeta(\eta)|_{\rho=0}\bigg(1 + \frac{\rho^2}{H^2}c(i\mu,\eta)\bigg)\,\,,
\end{equation}
where $P_\zeta(\eta)|_{\rho=0}$ is the power spectrum in absence of interactions 
(whose derivative goes to zero as $k^2\eta$ for $\eta\to 0^-$), 
and the function $c(i\mu,\eta)$ is defined by
\begin{equation}
\label{eq:app_D-D}
\begin{split}
&c(i\mu,\eta) = 2\pi\,\text{Re}\bigg[\int_x^{\infty}\dif x_1\,\frac{H_{i\mu}^{(1)}(x_1)}{\sqrt{x_1}}
\int_{x_1}^{\infty}\dif x_2\,\frac{e^{ix_1}H^{(2)}_{i\mu}(x_2)e^{-ix_2}}{\sqrt{x_2}}-\frac{e^{-ix_1}H^{(2)}_{i\mu}(x_2)e^{-ix_2}}{\sqrt{x_2}}\bigg] \\
&\hphantom{c(i\mu,\eta) } = -4\pi\,\text{Im}\bigg[\int_x^{\infty}\dif x_1\,\frac{H_{i\mu}^{(1)}(x_1)\sin x_1}{\sqrt{x_1}}
\int_{x_1}^{\infty}\dif x_2\,\frac{H^{(2)}_{i\mu}(x_2)e^{-ix_2}}{\sqrt{x_2}}\bigg]\,\,.
\end{split}
\end{equation}
The above integral is over $x=-k\eta$, 
the parameter $i\mu=\nu$ is defined by $\mu^2 = \frac{m^2}{H^2}-\frac{9}{4}$, 
and we dropped the $\pm i\epsilon$ from the upper limits for simplicity 
(they can be reinstated straightforwardly by requiring convergence in the infinite past). 

The correction to the power spectrum $c(i\mu,0)$ is computed in many works \cite{Chen:2009zp,Chen:2012ge,Lee:2016vti}: 
for our purposes it is sufficient to compute the derivative of \eq{app_D-D} with respect to $x$ and see how fast it goes to zero for $x\to 0$. 
Indeed, consider the two terms in \eq{app_D_intro-F} of the main text, 
\ie the contributions to the squeezed CFC bispectrum coming from the change in the time coordinate: 
we see that at order $(\rho/H)^2$ there will be a correction of the form\footnote{There will obviously be 
also the equivalent of these two terms with the $\mathcal{O}(\rho^2/H^2)$ 
corrections contained in the cross-spectra $P_{\zeta\xi^0}$ and $P_{\zeta B^0_{ij}}$. 
We do not consider them since they show a behavior similar to the one we discuss in the rest of the appendix.}
\begin{equation}
\label{eq:app_D_middle-A}
P_{\zeta\xi^0}(\eta,k_\ell)P_\zeta(k_s)\frac{\dif c}{\dif\eta}\,\,,\quad P_{\zeta B^0_{ij}}(\eta,k_\ell)P_\zeta(k_s)\frac{\dif c}{\dif\eta}\,\,.
\end{equation}
We write the derivative as
\begin{equation}
\label{eq:app_D-E}
\frac{\dif c}{\dif x} = 4\pi\,\text{Im}\bigg[\frac{H_{i\mu}^{(1)}(x) \sin x }{\sqrt{x}}\int_{x}^{\infty}\dif y\,\frac{H^{(2)}_{i\mu}(y)e^{-iy}}{\sqrt{y}}\bigg]\,\,,
\end{equation}
and we start by considering the case $\mu\geq 0$. 
The integral in \eq{app_D-E} can be carried out analytically, \ie
\begin{equation}
\label{eq:app_D-F}
\begin{split}
&I(x)=\int\dif x\,\frac{H^{(2)}_{i\mu}(x)e^{-ix}}{\sqrt{x}} \\
&\hphantom{I(x) } = \frac{2e^{-\frac{\pi\mu}{2}}\sqrt{x}}{\pi}\bigg[
\frac{e^{-\frac{\pi\mu}{2}}2^{-i\mu}x^{i\mu}\Gamma(-i\mu)\,{{}_2F_2}\big(\frac{1}{2}+i\mu,\frac{1}{2}+i\mu;\frac{3}{2}+i\mu,1+2i\mu;-2ix\big)}{2\mu-i} \\
&\hphantom{I(x) = \frac{2e^{-\frac{\pi\mu}{2}}\sqrt{x}}{\pi}\bigg[ } 
- \frac{e^{\frac{\pi\mu}{2}}2^{i\mu}x^{-i\mu}\Gamma(i\mu)\,{{}_2F_2}\big(\frac{1}{2}-i\mu,\frac{1}{2}-i\mu;\frac{3}{2}-i\mu,1-2i\mu;-2ix\big)}{2\mu+i}\bigg]\,\,.
\end{split}
\end{equation}
The value of $I(\infty)$ can be computed via an asymptotic expansion of ${{}_2F_2}$, following \cite{Chen:2012ge}. 
More precisely, we have
\begin{equation}
\label{eq:app_D_middle-B}
\begin{split}
&{{}_2F_2}(a,a;b_1,b_2;z)\sim\frac{\Gamma(b_1)\Gamma(b_2)}{\Gamma(a)^2} e^zz^{2a-b_1-b_2} \\
&\hphantom{{{}_2F_2}(a,a;b_1,b_2;z)\sim } 
+ \frac{\Gamma(b_1)\Gamma(b_2)}{\Gamma(a)\Gamma(b_1-a)\Gamma(b_2-a)}(-z)^{-a}\big[\log(-z)-\psi(b_1-a) \\
&\hphantom{{{}_2F_2}(a,a;b_1,b_2;z)\sim 
+ \frac{\Gamma(b_1)\Gamma(b_2)}{\Gamma(a)\Gamma(b_1-a)\Gamma(b_2-a)}(-z)^{-a}\big[ } 
- \psi(b_2-a)-\psi(a)-2\gamma\big]\,\,,
\end{split}
\end{equation}
where $\psi(z)$ is the Digamma function. 
The first term on the right-hand side of \eq{app_D_middle-B} can be neglected: 
indeed it goes to zero exponentially at large $x$ thanks to the $i\epsilon$ rotation in the complex plane. 
The final result is
\begin{equation}
\label{eq:app_D_middle-C}
I(\infty) = \frac{\sqrt{\pi}\,(1+i)}{2}\frac{e^{\frac{\pi\mu}{2}}(1+e^{-2\pi\mu})}{\cos^2(i\pi\mu)}\,\,.
\end{equation}
Besides, since ${{}_2F_2}$ is analytic in $x=0$, 
we see that $I(x)$ goes to zero as $\sqrt{x}\,x^{\pm i\mu}$ for small $x$. 
Therefore, 
the leading contribution to $\frac{\dif c}{\dif x}$ for $x\to 0$ comes from
\begin{equation}
\label{eq:app_D-G}
\frac{\dif c}{\dif x}\sim I(\infty)\frac{H^{(1)}_{i\mu}(x)\sin x}{\sqrt{x}}\,\,.
\end{equation}
The Hankel function behaves as $H^{(1)}_{i\mu}(x)\sim x^{\pm i\mu}$ for small $x$, 
therefore $\frac{\dif c}{\dif x}\sim\sqrt{x}\,x^{\pm i\mu}$, 
which goes to zero as $\sqrt{x}$ for $x\to 0$. 
We then see how, for $m\geq{3H}/{2}$, 
the two terms of \eq{app_D_middle-A} go to at late times. 
Indeed, we can compute $P_{\zeta\xi^0}$ and $P_{\zeta B^0_{ij}}$ using the solution for $\zeta$ with $\rho=0$. 
To do this, we need the expressions for $\xi^0$ and $B^0_{ij}$: 
using the results of \cite{Cabass:2016cgp} we see that 
the uniform shift $\xi^0(\eta,\vec{0})$ is equal to 
\begin{equation}
\label{eq:app_D_intro-C}
\xi^0(\eta,\vec{0}) = \int_0^\eta\dif s\bigg[(a_F/a)(s,\vec{0}) - \frac{\partial_0\zeta(s,\vec{0})}{\cH}\bigg]\,\,,
\end{equation}
with
\begin{subequations}
\label{eq:app_D_intro-D}
\begin{align}
&(a_F/a)(\eta,\vec{0}) = \int_0^\eta\dif s\bigg[\partial_0\zeta(s,\vec{0}) + \frac{\partial_i V^i(s,\vec{0})}{3}\bigg]\,\,, \label{eq:app_D_intro-D-1} \\
&V^i = \partial_i\digamma\,\,,\quad\digamma = 
{-e^{-\int_0^\tau\dif s\,\cH(s)}}\int_0^\eta\dif s\,\frac{e^{\int_0^s\dif u\,\cH(u)}\partial_0\zeta(s,\vec{0})}{\cH(s)}\,\,. \label{eq:app_D_intro-D-2} 
\end{align}
\end{subequations}
$B^0_{ij}$, instead, contains terms of the form 
\begin{equation}
\label{eq:app_D_middle-D}
B^0_{ij}\supset\bigg\{\partial_0\zeta,{-\frac{\zeta}{\cH}},\partial_i V^i\bigg\}\,\,,
\end{equation}
where $V^i$ is defined in \eq{app_D_intro-D-2} above. 
The calculation of $P_{\zeta\xi^0}$ and $P_{\zeta B^0_{ij}}$ for $\rho=0$ is straightforward: their late-time behavior is 
\begin{equation}
\label{eq:app_D_middle-E}
P_{\zeta\xi^0}\sim\eta^2\,\,,\quad P_{\zeta B^0_{ij}}\sim\eta\,\,.
\end{equation}

As we take the mass of $\sigma$ to be less than $\frac{3H}{2}$, 
it is useful to switch to $\nu = i\mu$: 
$\nu = \frac{3}{2}$ corresponds to $m=0$, 
while $\nu = 0$ corresponds to $m = \frac{3H}{2}$. 
For $\nu < \frac{1}{2}$, 
we see from \eq{app_D-F} that $I(x)$ continues to go to zero for $x\to 0$. 
Therefore, the leading contribution to $\frac{\dif c}{\dif x}$ will still come from \eq{app_D-G}: 
the leading behavior, though, is changed to $\frac{\dif c}{\dif x}\sim x^{\frac{1}{2}\pm\nu}$. 
As the mass becomes equal to $\sqrt{2}H$, 
we cannot apply the same argument anymore, 
since both $I(x)$ and $I(\infty)$ have a pole at $\nu = \frac{1}{2}$. 
However, the difference $I(\infty)-I(x)$ is finite as $\nu\to\frac{1}{2}$,\footnote{This can be checked 
by using the Laurent series for \eqsII{app_D-F}{app_D_middle-C} around $\nu = \frac{1}{2}$.} 
and we still have $\frac{\dif c}{\dif x}\sim x^{\frac{1}{2}\pm\nu}$ for small $x$. 
Combining this result with \eq{app_D_middle-E}, 
we see that as long as $\nu<\frac{3}{2}$, 
the contribution coming from the change of the time coordinate vanishes at late times. 

As $\nu\to\frac{3}{2}$, instead, 
our results seem to imply that there is a finite and non-zero contribution from the coordinate change to CFC. 
However, this limit corresponds to taking the field $\sigma$ to be massless: 
therefore, it does not decay on super-Hubble scales (but goes to a constant), 
and the transfer of $\sigma$ to $\pic$ continues indefinitely after Hubble exit. 
Then, evaluating the power spectrum at $\eta\to 0^-$ is not justified. 
This can be seen by taking the $\nu\to\frac{3}{2}$ limit of \eq{app_D-C}, 
as done in \cite{Chen:2012ge}: 
the function $c(i\mu,0)$ has a singularity there, 
which means that the super-Hubble power spectrum diverges. 
To give a rough idea of how this behavior arises, 
we compute the leading correction in $\rho$ to the classical mode functions of $\pic$ for the action of \eq{app_D-A}. 
This does not allow to compute the full power spectrum 
(for which the full in-in calculation is needed), 
but will allow us to check the time dependence of $\zeta$ on super-Hubble scales. 
The equation of motion for $\pic$ is 
\begin{equation}
\label{eq:app_D-H}
\pic'' + 2\cH\pic + k^2\pic + \frac{\rho}{H}(3\cH^2\sigma+\cH\sigma')=0\,\,,
\end{equation}
where $\cH = aH = -\eta^{-1}$. 
The solutions for $\pic$ and $\sigma$ at zeroth order in $\frac{\rho}{H}$ are\footnote{Notice that the 
constants $\alpha$ and $\beta$ have dimension $\frac{3}{2}$ 
(as the creation and annihilation operators do at the quantum level).} 
\begin{subequations}
\label{eq:app_D-I}
\begin{align}
&\pic = \alpha_{\pi}\,\underbrace{\frac{iH}{\sqrt{2k^3}}(1-ix)e^{ix}}_{\hphantom{\pi_{(1)}\,}\equiv\,\pi_{(1)}} 
\;+\; \beta_{\pi}\,\underbrace{\bigg({-\frac{iH}{\sqrt{2k^3}}}\bigg)(1+ix)e^{-ix}}_{\hphantom{\pi_{(2)}\,}\equiv\,\pi_{(2)}}\,\,, \label{eq:app_D-I-1} \\
&\sigma = \alpha_\sigma\,\underbrace{\frac{\sqrt{\pi}\,He^{-\frac{i\pi\nu}{2} - \frac{i\pi}{4}+i\pi}}{2k^\frac{3}{2}}x^\frac{3}{2}H^{(1)}_\nu(x)}_
{\hphantom{\sigma_{(1)}\,}\equiv\,\sigma_{(1)}} 
\;+\; \beta_\sigma\,\underbrace{\frac{\sqrt{\pi}\,He^{\frac{i\pi\nu}{2} + \frac{i\pi}{4}+i\pi}}{2k^\frac{3}{2}}x^\frac{3}{2}H^{(2)}_\nu(x)}_
{\hphantom{\sigma_{(2)}\,}\equiv\,\sigma_{(2)}}\,\,, \label{eq:app_D-I-2}
\end{align}
\end{subequations}
where the Bunch-Davies initial conditions select $\alpha_\pi = 1$, $\beta_\pi = 0$ and $\alpha_\pi = 1$, $\beta_\sigma=0$. 
We can then solve \eq{app_D-H} up to first order in $\frac{\rho}{H}$ as
\begin{equation}
\label{eq:app_D-J}
\begin{split}
&\pic(\eta) = \pi_{(1)}(\eta) 
+ \frac{\rho}{H}\bigg[{-\pi_{(1)}(\eta)\int^\eta\dif s\,\frac{\pi_{(2)}(s)J(s)}{w(s)}} 
+ \pi_{(2)}(\eta)\int^\eta\dif s\,\frac{\pi_{(1)}(s)J(s)}{w(s)}\bigg]\,\,,
\end{split}
\end{equation}
where we defined $J\equiv 3\cH^2\sigma+\cH\sigma'$ 
(that is evaluated using the mode functions of \eq{app_D-I}, 
since any correction will be higher order in $\frac{\rho}{H}$), 
$w\equiv\pi_{(1)}\pi'_{(2)} - \pi_{(2)}\pi'_{(1)} = iH^2\eta^2$, 
and we have denoted by $\int^\eta\dif s$ the fact that we do not choose a particular initial time to compute the inhomogeneous solution. 
By Taylor expanding the mode functions for $\pi_\text{c}$, 
and using the fact that $\pi_{(1)}(\eta)+\pi_{(2)}(\eta) = \mathcal{O}(\eta^3)$ for $\eta\to 0^{-}$, 
we see that the leading contribution from the $\pi$-$\sigma$ exchange at late times is given by 
\begin{equation}
\label{eq:app_D-K}
\pic(\eta)\supset\frac{\rho}{H}\frac{i H^2}{3}\bigg[\eta^3\int^\eta\dif s\,\frac{J(s)}{w(s)} + \int^\eta\dif s\,\frac{s^3J(s)}{w(s)}\bigg]\,\,,
\end{equation}
where $J(\eta)$ scales as $(-\eta)^{-\frac{1}{2}\pm\nu}$ at late times. 
This result shows that for $\nu=\frac{3}{2}$, 
the time derivative of the mode function for small $\eta$ goes as $-\eta^{-1}$, 
so that the power spectrum will diverge logarithmically on super-Hubble scales.





\bibliographystyle{utphys}
\bibliography{refs}

\end{document}